\documentclass[journal,10pt]{IEEEtran}

\usepackage{theorem}   
\usepackage{verbatim}
\usepackage{amssymb}  
\usepackage{color}
\usepackage{bm}
\usepackage{xcolor}
\usepackage{algorithm}
\usepackage{mathrsfs}
\usepackage{graphicx}
\usepackage{float}
\usepackage{subfigure}
\newtheorem{theorem}{Theorem}
\newtheorem{proposition}{Proposition}

\newtheorem{lemma}{Lemma}

\newtheorem{T-Prob}{Transformed Problem}

\usepackage{mathrsfs}
\usepackage{cite}
\usepackage{multirow}
\usepackage{amsmath,amssymb,amsfonts}

\usepackage{algpseudocode}
\usepackage{algorithmicx}

\usepackage{array}
\usepackage{stfloats}
\usepackage{url}

\usepackage{easyReview}

\begin{document}

\title{A Manifold Learning-based CSI Feedback Framework for FDD Massive MIMO}
\author{Yandi Cao, Haifan Yin,~\IEEEmembership{Senior Member,~IEEE}, Ziao Qin, Weidong Li, Weimin Wu and M{\'e}rouane Debbah,~\IEEEmembership{Fellow,~IEEE}
\thanks{Y. Cao, H. Yin, Z. Qin, W. Li and W. Wu are with School of Electronic Information and Communications, Huazhong University of Science and Technology, 430074 Wuhan, China (e-mail: ydcao@hust.edu.cn, yin@hust.edu.cn, ziao\_qin@hust.edu.cn, weidongli@hust.edu.cn, wuwm@hust.edu.cn).}
\thanks{M. Debbah is with KU 6G Research Center, Khalifa University of Science and Technology, P O Box 127788, Abu Dhabi, UAE (email: merouane.debbah@ku.ac.ae) and also with CentraleSupelec, University Paris-Saclay, 91192 Gif-sur-Yvette, France.}
\thanks{The corresponding author is Weimin Wu.}
\thanks{A part of this work \cite{cao2022manifold} was presented in the IEEE International Conference on Communications (IEEE ICC 2022).}
\thanks{This work was supported by the National Natural Science Foundation of China under Grants 62071191, 62071192, and 1214110.}
}

\maketitle

\begin{abstract}
Massive multi-input multi-output (MIMO) in Frequency Division Duplex (FDD) mode suffers from heavy feedback overhead for Channel State Information (CSI). In this paper, a novel manifold learning-based CSI feedback framework (MLCF) is proposed to reduce the feedback and improve the spectral efficiency for FDD massive MIMO. Manifold learning (ML) is an effective method for dimensionality reduction. However, most ML algorithms focus only on data compression, and lack the corresponding recovery methods. Moreover, the computational complexity is high when dealing with incremental data. Considering to utilize the intrinsic manifold structure where the CSI samples reside, we propose a landmark selection algorithm to describe the topological skeleton of this manifold. Based on the learned skeleton, the local patch of the incremental CSI on the manifold can be easily determined by its nearest landmarks. This motivates us to propose an incremental CSI compression and reconstruction scheme by keeping the local geometric relationships with landmarks invariant. We theoretically prove the convergence of the proposed landmark selection algorithm. Meanwhile, the upper bound on the error of approximating CSI with landmarks is derived. Simulation results under an industrial channel model of 3GPP demonstrate that the proposed MLCF outperforms existing deep learning based algorithms.
\end{abstract}


\IEEEpeerreviewmaketitle

\section{Introduction}

\IEEEPARstart{M}{assive} multiple-input multiple-output (MIMO) is one of the key enabling technologies for the fifth generation (5G) wireless communication systems \cite{Larsson2014Massive} \cite{Marzetta2015massive}. By deploying a large number of antennas at the base station (BS), massive MIMO systems have the potential to provide high spectral and energy efficiency \cite{Marzetta2010Noncooperative}. These performance gains depend on accurate and timely channel state information (CSI) at the transmitter. Since full channel reciprocity is not available in the frequency division duplex (FDD) system, the downlink CSI has to be estimated from pilots at the user equipment (UE) and then fed back to the BS \cite{love2008overview} \cite{qin2022partial}. Unfortunately, the dimension of the channel matrix scales with the number of BS antennas, exacerbating the feedback overhead in massive MIMO systems \cite{Liang2020Deep}. In addition, the amount of feedback is constrained by the coherence time and coherence bandwidth of the channel, both of which are limited in a mobility environment with multipath components. Therefore, one of the most challenging tasks for FDD massive MIMO is how to reduce the CSI feedback overhead while keeping the accuracy of the reconstructed CSI at the BS high.

The urgent demand for limited CSI feedback has motivated extensive research. A vector quantization codebook \cite{Raghavan2007Systematic} \cite{kim2012differential} is designed to be shared between the BS and the UE, with the UE feeding back the index of codeword that best matches the downlink CSI. The size of the codebook required to maintain the level of communication quality grows exponentially with the number of antennas, which is costly in the case of massive MIMO. In view of the sparsity of massive MIMO channel in a certain transform domain, the theory of compressed sensing (CS) is introduced in \cite{kuo2012Compressive, gao2018compressive,alevizos2018limited, sun2018limited}. Such approaches rely on the sparse assumption of the channel, which may not always be present in practice. Moreover, iterative algorithms are proposed for CSI recovery, resulting in heavy computational costs. In particular, in order to exploit the sparsity of the channel, some works \cite{alevizos2018limited}\cite{sun2018limited} parameterize the azimuth angle of arrival (AoA) and azimuth angle of departure (AoD), and design over-complete dictionaries to approximate the steering vectors. The resolution of AoA and AoD is reflected in the size of the dictionary, which directly affects the reconstruction performance.

Due to the powerful feature extraction ability, deep learning (DL) has recently shown great potential in the field of wireless communication, such as hybrid precoding \cite{tian2020randomized}, channel prediction \cite{yang2019deep} and channel extrapolation \cite{zhang2021deep} \cite{lin2021deep}. This has also inspired a number of DL-based algorithms to address the problem of limited CSI feedback in FDD massive MIMO. An auto-encoder network named CsiNet \cite{Chao-Kai2018Deep} has been proposed for CSI compression and recovery. Specifically, an encoder compresses the channel matrices into codewords at the UE, and then a decoder reconstructs the channel matrices from the received codewords at the BS. Enlightened by the fact that DL is an appealing way for CSI feedback, a multi-resolution CRBlock \cite{Zhilin2020Multi} is designed to improve the robustness under various scenarios and compression ratios. To extract the temporal correlation of CSI in adjacent time slots, the authors in \cite{Wang2018Deep} design a long short-term memory (LSTM) network to further improve the reconstruction quality. In \cite{mashhadi2020distributed}, a fully-convolutional neural network called DeepCMC is proposed that jointly considers the CSI compression, quantization, and entropy coding to enable a trade-off between the CSI reconstruction quality and feedback overhead. Nevertheless, the DL-based algorithms typically require very large training data sets. Other relevant works have also been carried out to avoid the use of DL. The authors in \cite{nerini2022machine} employ Principal Component Analysis (PCA), a classical Machine Learning technique, to compress the CSI into the latent space dimension. 

The aforementioned research mainly focused on the characteristics of the channel structure in the angular-delay/frequency domain \cite{Chao-Kai2018Deep}\cite{li2022multi} or the structure of the channel covariance matrix \cite{nerini2022machine}\cite{yin2013coordinated}\cite{yin2014dealing}. However, the intrinsic manifold structure where the CSI resides is neglected. It has been verified in \cite{love2003grassmannian}\cite{bhogi2020learning} that connecting manifold to codebook design is possible under the assumption that the optimal beamformer for each channel resides on the Grassmann manifold. Similarly, we consider reformulating the limited CSI feedback problem as a manifold optimization problem by means of manifold learning (ML) \cite{Tong2008Riemannian, Roweis2000Nonlinear, Zhenyue2004Principal}. Even though the manifold structure is taken into consideration in both our work and that of \cite{bhogi2020learning}, there are fundamental differences in the problems  solved and the algorithms designed. The authors in \cite{bhogi2020learning} reduce the beamforming codebook design problem to a clustering problem on the Grassmann manifold where the cluster centroids form the codebook. In this paper, however, we deal with the CSI compression and reconstruction problem by maintaining the local geometric structure of the manifold unchanged before and after compression.

As a nonlinear dimensionality reduction approach, ML can map a high-dimensional data set into a low-dimensional space while preserving the intrinsic manifold structure in the data. However, two open issues hinder ML from being applied in practice. Firstly, most ML algorithms, such as Locally Linear Embedding (LLE) \cite{Roweis2000Nonlinear} and Local Tangent Space Alignment (LTSA) \cite{Zhenyue2004Principal}, work in a “batch” mode, which means that dimensionality reduction cannot be performed in an incremental way. When a new sample arrives, the original data set needs to be updated to incorporate the new sample, and the ML algorithms have to be rerun with the new data set. This process undoubtedly increases the computational cost. Secondly, most ML algorithms lack an inverse mapping from the low-dimensional embedding to the high-dimensional data, i.e., they focus on data compression, while the corresponding reconstruction methods are not provided. Some works \cite{Zhenyue2004Principal} \cite{zhang2004reconstruction} attempt to recover the original data by introducing kernel regression functions to fit the high-dimensional data. However, the number of required kernel functions scales linearly with the dimension of high-dimensional data, so determining the parameters of these functions might be costly when the data is of large dimensionality.

In order to address the aforementioned issues in traditional ML algorithms and adopt the idea of ML to limited CSI feedback, we propose a novel ML-based CSI feedback framework (MLCF) to improve the spectral efficiency (SE) of FDD massive MIMO systems. It is assumed that the CSI samples lie on a smooth, low-dimensional manifold embedded in a high-dimensional space. Without prior knowledge of the intrinsic manifold, a data set consisting of high-density CSI samples is constructed to characterize the manifold. It may contain spatially redundant samples, thus we carefully select representative landmarks from the data set to describe the manifold skeleton. At this point, for the newly sampled CSI, its local patch on the manifold can be easily identified by its nearest neighbors in the landmarks rather than the data set. Then the local geometric relationship between CSI and the landmarks in the high-dimensional space can be easily determined. This relationship is expected to be maintained between the low-dimensional embedding of CSI and the landmarks in low-dimensional space, so as to calculate the compressed CSI. This idea also allows for the reconstruction of the original CSI.

The main contributions of this paper are summarized as follows:
\begin{enumerate}
\item To the best of our knowledge, this paper is the first to employ the idea of ML to the limited CSI feedback problem. Without having prior knowledge of the manifold structure, we design landmarks to characterize the topological skeleton of the manifold where the CSI samples reside. An alternating iteration optimization algorithm is proposed to select representative landmarks from the manifold. Additionally, the closed-form solutions of the algorithm are provided.
\item Based on the manifold skeleton, the local patch of the incremental CSI on the manifold can be determined by its nearest neighbors in landmarks. Inspired by this, we propose an incremental CSI compression and reconstruction scheme that alleviates the drawbacks of traditional ML algorithms, including the high complexity when handling the incremental data and the absence of inverse reconstruction. 
\item We prove the convergence of the proposed alternating iterative optimization algorithm, and show that the value of the objective function decreases monotonically in each iteration. The theoretical proof and the simulation results under an industrial channel model of 3GPP are in agreement.
\item We derive an upper bound on the error of approximating the CSI with landmarks. We show the main factors that affect the error are the number and correctness of the nearest neighbors in the landmarks. If all neighbors lie in a sufficiently compact region, the approximation error is quite small.
\end{enumerate}

Simulation results under the industrial channel model of 3GPP demonstrate large gains in terms of the normalized mean square error (NMSE) of the reconstructed CSI. In particular, when the compression ratio is 1/4, the proposed MLCF method brings a gain of at least 7 dB compared with existing DL-based methods. 

Notations: We use boldface to denote vectors and matrices. $({\mathbf{A}})^T$ and $({\mathbf{A}})^H$ denote the transpose and conjugate transpose of the matrix ${\mathbf{A}}$, respectively. ${\mathop{ \Re}\nolimits} \{ \cdot \}$ and ${\mathop{ \Im}\nolimits} \{ \cdot \}$  represent the real and imaginary parts of a matrix, respectively. ${\| \cdot \|}_2$ is the Euclidean norm of a vector, and ${\| \cdot \|}_F$ is the Frobenius norm of a matrix. $\rm{tr}\{\cdot\}$ denotes the trace of a square matrix. ${\mathbf{e}} \in {\mathbb{R}^{k \times 1}}$ is a column vector in which all elements are ones. ${\mathbf{A}_{i*}}$ is the $i$-th row of the matrix ${\mathbf{A}}$. ${{\rm span}}({\mathbf{A}})$ denotes the column space of ${\mathbf{A}}$. ${\rm{diag}}({q_1}, {q_2}, \ldots, {q_n})$ represents a diagonal matrix with ${q_1}, {q_2}, \ldots, {q_n}$ at the main diagonal. ${\mathbb{E}} \{ \cdot \}$ denotes the expectation.

\section{System Model}
In this work, we consider a single-cell FDD massive MIMO system, where the BS is equipped with a uniform linear array (ULA) with ${N_t}$ antennas and the UE is equipped with a single antenna. The antenna elements in ULA are separated horizontally by half a carrier wavelength. The system operates in orthogonal frequency division multiplexing (OFDM) modulation with a $\Delta f$ subcarrier spacing and ${N_f}$ subcarriers. For brevity of exposition, we focus on an arbitrary UE in the cell. The received signal at the UE is expressed as
\begin{equation}
    {r_i} = {\mathbf{h}_i} {\mathbf{p}_i} {s_i} + {n_i},
    \label{received signal}
\end{equation}
where ${\mathbf{h}_i} \in \mathbb{C}^{1 \times {N_t} }$, ${\mathbf{p}_i} \in \mathbb{C}^{{N_t} \times 1} $, ${s_i} \in \mathbb{C}$, and ${n_i} \in \mathbb{C}$ denote the downlink channel vector, the precoding vector, the transmit data symbol, and the additive noise at the $i$-th subcarrier, respectively. In the FDD system, the transmitter needs the knowledge of the accurate and instantaneous downlink CSI to design ${\mathbf{p}_i}$ to achieve high SE. 

The clustered delay line (CDL) channel model conforming to the 3GPP TR 38.901 specifications \cite{3GPP2019Study} is considered. There exists $N_c$ scattering clusters in the propagation environment, each containing $N_p$ rays. The downlink channel vector between the BS and the UE at the certain frequency $f_i$ and time $t$ is modeled as
\begin{equation}
    \begin{array}{l}
    {\mathbf{h}_{f_i}}(t) = \sum\limits_{n = 1}^{N_c} {\sum\limits_{m = 1}^{N_p} {\alpha _{n,m}}  {e^{ - j2\pi {f_i}{\tau _{n,m}}}} {e^{j {w_{n,m}}t}} {\textbf{a}(\theta_{n,m})}  } 
    \end{array},
    \label{CDL channel}
\end{equation}
where ${\alpha _{n,m}}$, ${\tau _{n,m}}$, ${w_{n,m}}$ and ${\theta_{n,m}}$ are the channel gain, the time delay, the Doppler frequency, and the AoD of the $m$-th ray in the $n$-th cluster, respectively. The steering vector ${\textbf{a}(\theta)}$ is formulated as
\begin{equation}
    {\textbf{a}(\theta)} = \left[ {\begin{array}{*{20}{c}}
1&{{e^{j2\pi \frac{{D\sin \theta }}{{{\lambda _0}}}}}}& \ldots &{{e^{j2\pi \frac{{(N_t-1)D\sin \theta }}{{{\lambda _0}}}}}}
\end{array}} \right] \in {\mathbb{C}^{1 \times {N_t}}},
\end{equation}
with ${D}$ and ${\lambda _0}$ being the spacing between element antennas and the central carrier wavelength, respectively.

The channel vectors over $N_f$ subcarriers are integrated into a wideband channel matrix ${\mathbf{H}} \in {\mathbb{C}^{{N_f} \times {N_t}}}$, defined as
\begin{equation}
    {\mathbf{H}}(t) = \left[ {\begin{array}{*{20}{c}}
    {\mathbf{h}_{f_1}^T}(t) & {\mathbf{h}_{f_2}^T}(t) & \ldots &{\mathbf{h}_{f_{N_f}}^T}(t)
    \end{array}} \right]^T .
    \label{channel matrix}
\end{equation}
The total number of feedback parameters is proportional to $N_f$ or ${N_t}$. Thus, how to compress CSI efficiently becomes a challenging task for FDD massive MIMO. 

Given that ML only supports the calculation of real values, the complex channel matrix ${\mathbf{H}}$ is decomposed into real and imaginary parts, stacked as below
\begin{equation}
    \widetilde {\mathbf{H}}(t) = \left[ {\begin{array}{*{20}{c}}
    {{\mathop{\Re}\nolimits} \left\{ {\mathbf{H}}(t) \right\}}\\
    {{\mathop{\Im}
    \nolimits} \left\{ {\mathbf{H}}(t) \right\}}
    \end{array}} \right] \in {\mathbb{R}^{2{N_f} \times {N_t}}} .
    \label{channel decomposition}
\end{equation}
Note that here we consider perfect channel samples. However, our method still works well in the case of noisy channel estimation, as will be shown in Sec. \ref{Sec:Sim}. Suppose that the real CSI is sampled from a high-dimensional manifold $\mathcal{M}$. The manifold topology can be characterized by a data set consisting of high-density samples on the manifold. However, the large-sized data set may have spatially redundant samples leading to high computational complexity. Therefore, we replace the data set by selecting representative landmarks that not only have small size but also can characterize the topological skeleton of $\mathcal{M}$ well. 

In details, by means of ML, the UE compresses the downlink CSI estimated from the pilot signals as
\begin{equation}
   \label{compression equation}
    \mathbf{Y} =  f(\widetilde {\mathbf{H}}, \mathbf{D}_{\rm{H}}^{\rm{dr}},\mathbf{D}_{\rm{L}}^{\rm{dr}}),
\end{equation}
where $\mathbf{Y} \in {\mathbb{R}^{d \times N_t}}$ is the low-dimensional embedding of $\widetilde {\mathbf{H}}$, $f(\cdot)$ denotes the compression function, $\mathbf{D}_{\rm{H}}^{\rm{dr}}$ and $\mathbf{D}_{\rm{L}}^{\rm{dr}}$ are a set of dictionaries for dimensionality reduction. The two dictionaries consist of selected landmarks on the high-dimensional and low-dimensional manifolds, respectively. After receiving $\mathbf{Y}$, the reconstruction operation is performed at the BS to recover the downlink CSI by
\begin{equation}
    \mathbf{\hat{H}} = g(\mathbf{Y}, \mathbf{D}_{\rm{H}}^{\rm{rc}},\mathbf{D}_{\rm{L}}^{\rm{rc}}),
\end{equation}
where $g(\cdot)$ is the reverse operation of compression, $\mathbf{D}_{\rm{H}}^{\rm{rc}}$ and $\mathbf{D}_{\rm{L}}^{\rm{rc}}$ are a set of reconstruction dictionaries. Note that these two sets of dictionaries are learned in advance.

\section{Proposed Landmark Selection Method for Manifold Learning}
In this section, we first describe how to obtain the intrinsic manifold structure where the CSI resides. Then, a landmark selection  algorithm is proposed to characterize the topological skeleton of this manifold. Meanwhile, the closed-form solutions and convergence analysis of the algorithm are provided.

\subsection{Manifold structure of CSI}
\label{subsectionA}
In real-world systems, it is common to deal with numerous high-dimensional data with low intrinsic degrees of freedom. Frustratingly, the high-dimensional data suffers from the curse of dimensionality \cite{Tong2008Riemannian} \cite{ van2009dimensionality}, exacerbating storage requirements and computational burden. A typical strategy to alleviate this problem is ML, which aims to map a high-dimensional data set into a low-dimensional space while maintaining the underlying structure in the data. 

To satisfy the ML assumption, we consider all CSI samples residing on a $d$-dimensional manifold $\mathcal{S}$ embedded in a $2{N_f}$-dimensional space $\mathcal{M}$. Since the true manifold distribution $\mathcal{M}$ is not accessible, we collect a large number of CSI samples from each UE to form a training set in the hope that it is close to the original manifold structure. In FDD massive MIMO systems, a total of ${T_s}$ downlink CSI samples are observed to form the data set
\begin{equation}
    {\mathbf{X}} = \left[ {\begin{array}{*{20}{c}}
     \widetilde {\mathbf{H}}(1) & \ldots & \widetilde {\mathbf{H}}(T_s)
    \end{array}} \right] ,
    \label{high-dimensional dataset}
\end{equation}
which serves to represent the high-dimensional space $\mathcal{M}$. For simplicity, let ${{\mathbf{x}}_i}$ be the $i$-th column of ${\mathbf{X}} \in {\mathbb{R}^{2{N_f} \times N}}$, and $N = {N_t}{T_s}$ be the size of ${\mathbf{X}}$. The process of dimensionality reduction is defined as
\begin{equation}
    {{\mathbf{y}}_i} = f\left( {{\mathbf{x}}_i} \right) + {{\boldsymbol{\epsilon}}_i} ,
    \label{emdedded mapping}
\end{equation}
where ${f(\cdot): \mathcal{M} \rightarrow \mathcal{S}}$ is the compression function, ${{\boldsymbol{\epsilon}}_i}\in {\mathbb{R}^{d \times 1}}$ is the error. Then the corresponding low-dimensional embedding of ${\mathbf{X}}$ is obtained by ${\mathbf{Y}} = f({\mathbf{X}}) \in {\mathbb{R}^{d \times N}}$, which can characterize the low-dimensional manifold $\mathcal{S}$. Obviously, the compression ratio is
\begin{equation}
    \gamma = \frac{dN} {2{N_f}N} = \frac{d} {2{N_f}} ,
    \label{compression ratio}
\end{equation}
where $d \ll 2{N_f}$. 

In short, given a high-dimensional data set ${\mathbf{X}}$, the task of ML is to obtain its low-dimensional representation ${\mathbf{Y}}$ while finding out the mapping function $f$. However, data reconstruction, as the inverse process of dimensionality reduction, is frequently overlooked, resulting in difficultly recovering the original data from the low-dimensional embedding. In the following, we will discuss how to obtain the reconstruction mapping ${g(\cdot): \mathcal{S} \rightarrow \mathcal{M}}$. Once the explicit functions $f$ and $g$ are known, the newly sampled CSI can be efficiently compressed and reconstructed.

\subsection{Landmark selection}
\begin{figure}[htbp]
\centering
\begin{minipage}[t]{0.48\linewidth}
\centering
\includegraphics[width=4cm]{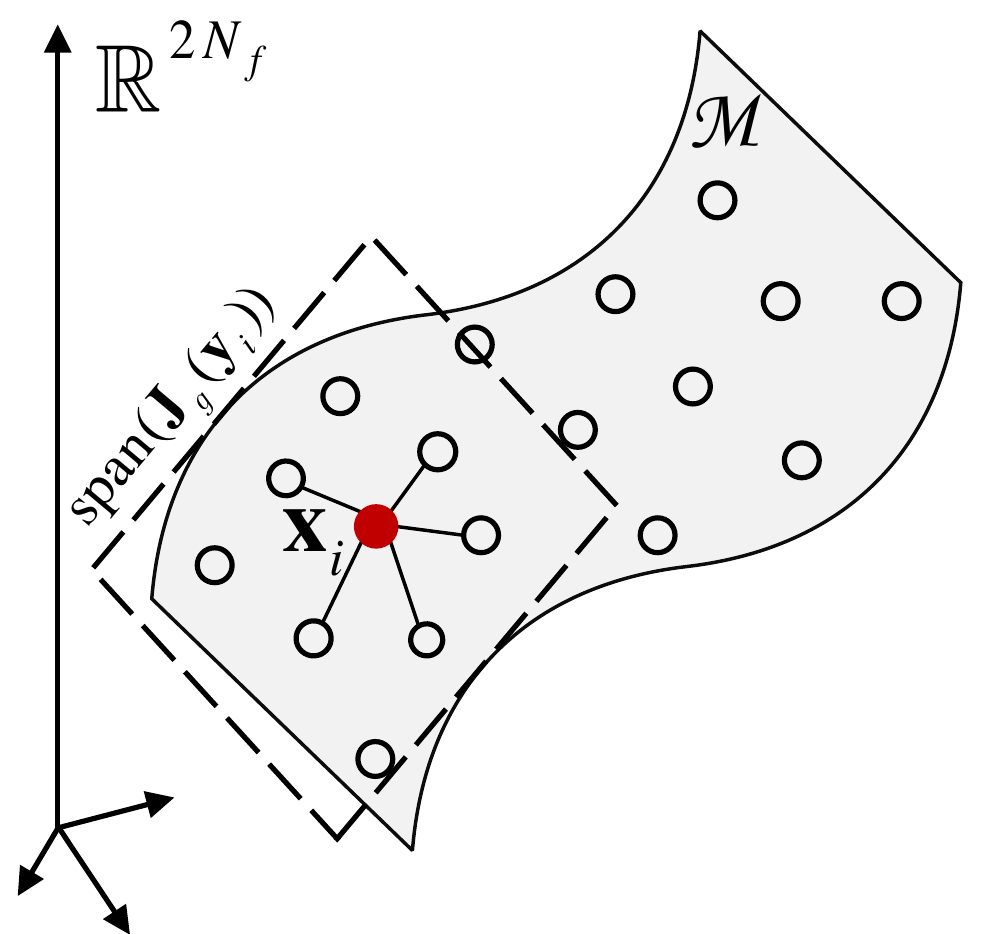}
\centerline{(a)}
\end{minipage}
\begin{minipage}[t]{0.48\linewidth}
\centering
\includegraphics[width=4cm]{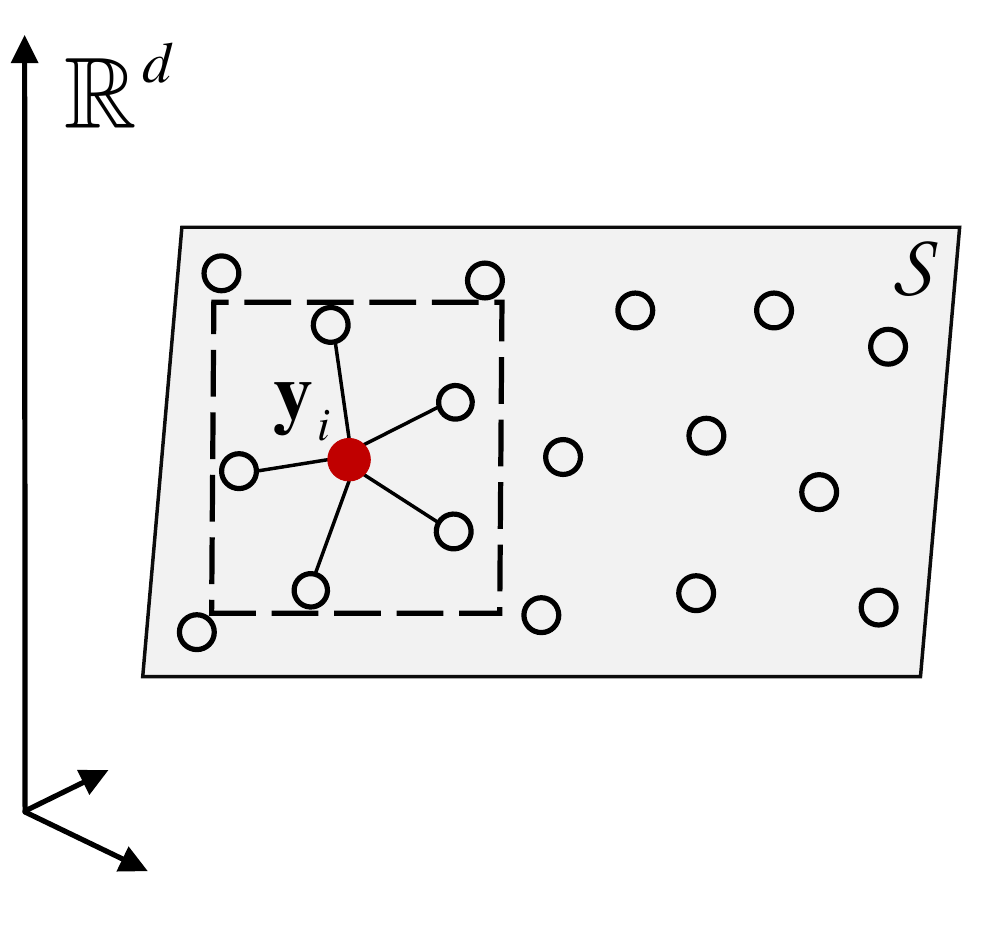}
\centerline{(b)}
\end{minipage}
\caption{An illustration of the high-dimensional manifold (a) and the corresponding low-dimensional embedding (b). The CSI data ${{\mathbf{x}}_i}$ (the solid circle) is sampled from the surface in (a). ${{\rm span}}({\mathbf{J}}_g({{\mathbf{y}}_i}))$ is the tangent space at ${{\mathbf{x}}_i}$. The hollow circles are the landmarks selected by our proposed Algorithm~\ref{alg1}. The local geometric relationship between ${{\mathbf{x}}_i}$ and its nearest landmarks remains unchanged before and after dimensionality reduction.}
\label{fig1}
\end{figure}

It should be emphasised that the complexity of most ML algorithms is strongly influenced by the dimensionality and quantity of data samples. Typically, the low-dimensional embedding is obtained by performing an eigenvector analysis on the data similarity matrix, whose size is $N \times N$. As a result, ML is not suitable for large-scale data sets, which demand excessive computational and storage resources. Considerable efforts have been devoted to deal with the case of the large data set. The idea of perfectly approximating the manifold skeleton with a collection of landmarks \cite{zhang2010clustered} emerges, however selecting the right landmarks is essential. This also motivates us to employ landmarks to streamline the data set ${\mathbf{X}}$, thus alleviating the computing burden. The representative landmarks are selected from the historical CSI data set to characterize the topological skeleton of $\mathcal{M}$. The main selection principle is that the landmarks can linearly approximate all samples on the manifold with minimal error. For the newly sampled CSI, its local patch on the manifold can be easily determined with the known landmarks (see Fig.~\ref{fig1} for an illustration). 

The problem of dimensionality reduction focuses on finding the mapping $f$ from the high-dimensional space to the low-dimensional space. We may think of first establishing a local mapping relationship between the previously collected data sets ${\mathbf{X}}$ and ${\mathbf{Y}}$, and then extending the local mapping to the global mapping. LLE \cite{Roweis2000Nonlinear} expects each sample and its neighbors to lie on or be close to a local patch on the manifold. In this case, each sample can be approximated by the linear combination of its nearest neighbors to preserve the local geometric property. Similar to LLE, we hope each sample on the manifold $\mathcal{M}$ can be linearly approximated by its $k$ nearest landmarks to capture the local geometric relationship. 

Let ${\mathbf{D}}_{\rm{H}}^{\rm{dr}} = \left[ {\begin{array}{*{20}{c}} {{\mathbf{d}}_1}&{{\mathbf{d}}_2}& \ldots &{{\mathbf{d}}_M} \end{array}} \right] \in {\mathbb{R}^{2{N_f} \times M}}$ be a high-dimensional dictionary, composed of landmarks selected from the high-dimensional space $\mathcal{M}$, to represent the manifold skeleton. Specifically, ${{\mathbf{d}}_i}, i = 1,2,\ldots,M$, is the $i$-th column in ${\mathbf{D}}_{\rm{H}}^{\rm{dr}}$, and $M\left(M\ll N\right)$ is the size of ${\mathbf{D}}_{\rm{H}}^{\rm{dr}}$. Right now, we have
\[ {{\mathbf{x}}_i} \approx \sum\limits_{j = 1}^M {{w_{ji}}{{\mathbf{d}}_j}} = {\mathbf{D}}_{\rm{H}}^{\rm{dr}} {{\mathbf{w}}_i}, \]
subject to 
\begin{equation}
    \left\{ {\begin{array}{*{20}{c}} {{\mathbf{e}}^T {{\mathbf{w}}_i} = 1\ \ i = 1, \ldots ,N}\\ 
    {{w_{ji}} = 0\ \ \ {\rm{\ \ if\ }}{j} \notin \mathcal{N}_ {{\mathbf{x}}_i} } \ \ \ \ \end{array}} \right. ,
    \label{x_linear aproximation}
\end{equation}
where ${{\mathbf{w}}_i}={\left[ {\begin{array}{*{20}{c}} {{w_{1i}}}&{{w_{2i}}}& \ldots &{{w_{Mi}}} \end{array}} \right]^T} \in {\mathbb{R}^{M \times 1}}$ is a weight vector, and $\mathcal{N} _{{\mathbf{x}}_i}$ is an index set containing the column indices of the $k$ nearest landmarks of ${{\mathbf{x}}_i}$. The sum of the weights is enforced to be one such that the local geometry of the manifold is invariant in scaling, rotating, and shifting the coordinate system. If ${{\mathbf{d}}_j}$ is not in the neighborhood of ${{\mathbf{x}}_i}$, the weight $w_{ji}$ is set to zero to ensure the locality constraint. 

Meanwhile, the above local relationship is expected to be tenable between ${{\mathbf{y}}_i}$ and the low-dimensional dictionary ${\mathbf{D}}_{\rm{L}}^{\rm{dr}} = \left[ {\begin{array}{*{20}{c}}
f({{\mathbf{d}}_1})&f({{\mathbf{d}}_2})& \ldots &f( {{\mathbf{d}}_M}) \end{array}} \right] $ $\in {\mathbb{R}^{d \times M}}$, i.e.,
\begin{equation}
    {{\mathbf{y}}_i} \approx \sum\limits_
    {j = 1}^M {{w_{ji}}f({{\mathbf{d}}_j})} =  {\mathbf{D}}_{\rm{L}}^{\rm{dr}} {{\mathbf{w}}_i}, 
    \label{y_linear aproximation}
\end{equation}
where the weight vector ${{\mathbf{w}}_i}$ and the index set $\mathcal{N} _{{\mathbf{y}}_i}$ are the same as that in Eq.~\eqref{x_linear aproximation}. In the following, we will analyze whether the local linear approximation based on landmarks is feasible and what affects the approximation error.

\newenvironment{prof}{{\it Proof}}{\hfill $\square$ \par}
\begin{proposition}\label{prop1}
The linear approximation error satisfies
\begin{equation}
    \begin{aligned}
    {\left\| {\mathbf{x}_i} - \sum\limits_{j = 1}^M {w_{ji}}{\mathbf{d}_j}  \right\|} \le \ & \xi \left\| {\mathbf{J}_g}({{\mathbf{y}}_i}) \right\|_F  \sum\limits_{j = 1}^M  \left\| {\mathbf{b}_j} - {{\mathbf{y}}_i}\right\| \\ 
    & + {\xi_1}\left\| {\boldsymbol{\Psi}} \right\|_F \sum\limits_{j = 1}^M {{\left\| {\mathbf{b}_j} - {{\mathbf{y}}_i} \right\|}^2},     
    \end{aligned}
\end{equation}
where $\xi = \max {\{ {w_{ji}}\} }_j$ is the largest entry in ${{\mathbf{w}}_i}$, $\xi_1 = \sqrt{\frac{N_f}{2} }\xi$, ${\mathbf{J}_g}({{\mathbf{y}}_i})$ is the Jacobi matrix of $g$ at ${{\mathbf{y}}_i}$, ${{\mathbf{b}}_j} = f({\mathbf{d}}_j)$ is the low-dimensional embedding of ${\mathbf{d}}_j$, and \[{\boldsymbol{\Psi}} = \left[ {\begin{array}{*{20}{c}} {\boldsymbol{\Psi}^T_{g_1}}({{\mathbf{y}}_i})& {\boldsymbol{\Psi}^T_{g_2}}({{\mathbf{y}}_i})& \ldots &{\boldsymbol{\Psi}^T_{g_{2N_f}}}({{\mathbf{y}}_i}) \end{array}} \right]^T ,\] among which ${\mathbf{\Psi}_{g_l}}({{\mathbf{y}}_i})$ is the Hessian matrix of the $l$-th component function $g_l$ of $g$ at ${{\mathbf{y}}_i}$.

\begin{prof}
: \rm{Please refer to Appendix \ref{appendixA}.}
\end{prof}
\end{proposition}

According to the above analysis, selecting appropriate landmarks in a compact region is the key that straightway impacts the error. However, the neighbor set may contain the incorrect landmarks due to noise interference or outliers. In this case, the local patch spanned by the nearest landmarks cannot sufficiently reflect the local geometry on the manifold. This prompts us to add an extra term to exclude far away landmarks as outliers. 

In order to minimize the error of linearly approximating the CSI samples in the data set ${\mathbf{X}}$ with landmarks, the objective function with respect to ${\mathbf{D}}_{\rm{H}}^{\rm{dr}}$ and the weight matrix ${\mathbf{W}} = \left[ {\begin{array}{*{20}{c}} {{{\mathbf{w}}}_1}&{{{\mathbf{w}}}_2}& \ldots &{{{\mathbf{w}}}_{N}} \end{array}} \right] \in {\mathbb{R}^{M \times N}}$ is formulated as
\begin{equation}
    \label{dr_objective function}
    \begin{aligned} 
    \langle {\mathbf{D}}_{\rm{H}}^{\rm{dr}},{\mathbf{W}} \rangle 
    = & \mathop {\arg \min }\limits_{{\mathbf{D}}_{\rm{H}}^{\rm{dr}},{\mathbf{W}}}   
    {\left\| {\mathbf{X}} - {\mathbf{D}}_{\rm{H}}^{\rm{dr}} {\mathbf{W}} \right\|_F^2} \\ 
    & + \lambda \sum\limits_{i = 1}^N  {\left\| {{\mathbf{y}}_i} - \sum\limits_{j = 1}^M {{w_{ji}}f\left( {{\mathbf{d}}_j} \right)} \right\|_2^2} + \mu {\left\| {\mathbf{W}} \right\|_{2,1}}     
    \end{aligned},
\end{equation}
which has the same constraints as Eq.~\eqref{x_linear aproximation}. Moreover, $\lambda$ and $\mu$ are the regularization parameters that tune the approximation error and the weight error, and ${\left\| \cdot \right\|_{2,1}}$ is the $L_{2,1}$ norm of the matrix, which is defined as ${\left\| \mathbf{W} \right\|_{2,1}} = \sum\nolimits_{i = 1}^M  {\left\| \mathbf{w}_{i*} \right\|_2} $. Adding the extra term ${\left\| {\mathbf{W}} \right\|_{2,1}}$ is expected to exclude distant landmarks by penalizing small row weights in ${\mathbf{W}}$. The weight ${\left\| \mathbf{w}_{i*} \right\|_2}$ is not squared and thus ${\left\| \mathbf{W} \right\|_{2,1}}$ penalizes more for small weights than ${\left\| \mathbf{W} \right\|_2}$. Since the explicit function $f$ is not available for a learning task, minimizing the above objective function becomes impractical. Hence, Lemma \ref{lemma1} \cite{boothby2003introduction} is introduced to discard the unknown function $f$, thereby obtaining a tractable optimization problem. 

\begin{lemma}
    \label{lemma1}
    Let ${\mathbf{p}} \in \mathcal{U}_{\mathbf{p}}$ be an open set on the manifold $\mathcal{M}$ with respect to ${\mathbf{p}}$, such that $\forall {\mathbf{q}} \in \mathcal{U}_{\mathbf{p}}$, the line segment $\overline{\mathbf{pq}}$ remains in $\mathcal{U}_{\mathbf{p}}$. If $\left| {{{\partial {f^m}} \mathord{\left/ {\vphantom {{\partial {f^m}} {\partial {{\mathbf{q}}^n}}}} \right. \kern-\nulldelimiterspace} {\partial {{\mathbf{q}}^n}}}} \right| \le C$, $1 \le m \le d$, $1 \le n \le 2N_f$, then for $\forall {\mathbf{q}} \in \mathcal{U}_{\mathbf{p}}$, we have 
    \begin{eqnarray*}
        {\left\| {f({\mathbf{q}}) - f({\mathbf{p}})} \right\|^2} \le 2{N_f}d{C^2}{\left\| {{\mathbf{q}} - {\mathbf{p}}} \right\|^2} .
    \end{eqnarray*}
\end{lemma}

This lemma is a generalization of the mean value theorem \cite{boothby2003introduction}. It shows that as ${\mathbf{q}}$ lies in a small neighborhood of ${\mathbf{p}}$, there exists an upper bound of ${\left\| {f({\mathbf{q}}) - f({\mathbf{p}})} \right\|^2}$. Assuming the aforementioned conditions hold, the second term in Eq.~\eqref{dr_objective function} meets the following inequality
\begin{equation}
    \begin{aligned}
    & {\left\| {{\mathbf{y}}_i} - \sum\limits_{j = 1}^M {w_{ji}}f( {{\mathbf{d}}_j} ) \right\|^2} 
    \mathop = \limits^{(a)} {\left\| \sum\limits_{j = 1}^M { w_{ji}\left[ f({\mathbf{x}_i}) - f({\mathbf{d}_j}) \right] } \right\|^2}  \\
    & \mathop \le \limits^{(b)} M\sum\limits_{j = 1}^M {w_{ji}^2{{\left\| {f({{\mathbf{x}}_i}) - f({{\mathbf{d}}_j})} \right\|}^2}} 
    \mathop \le \limits^{(c)} \tau_1 \sum\limits_{j = 1}^M {w_{ji}^2 {\left\| {{{\mathbf{x}}_i} - {{\mathbf{d}}_j}} \right\|}^2}
    \end{aligned},
\end{equation}
where (a) applies the constraint that ${\mathbf{e}}^T {{\mathbf{w}}_i} = \sum\nolimits_{j = 1}^M {{w_{ji}}} = 1$, (b) is derived from the inequality of arithmetic and geometric means, and in (c) $\tau_1 = 2{N_f}Md{C^2}$ is a constant deduced from Lemma \ref{lemma1}.

Therefore, the optimization problem of dimensionality reduction can be reformulated as
\begin{equation}
    \label{final objective function}
    \begin{aligned}
     & \langle {\mathbf{D}}_{\rm{H}}^{\rm{dr}},{\mathbf{W}} \rangle = \mathop {\arg \min }\limits_{{\mathbf{D}}_{\rm{H}}^{\rm{dr}},{\mathbf{W}}}  {\left\| {\mathbf{X}} - {\mathbf{D}}_{\rm{H}}^{\rm{dr}} {\mathbf{W}} \right\|_F^2} \\
    & \ \ \ \ \ \ \ \ \ \ \ \ \ \ \ + \lambda \sum\limits_{i = 1}^N  \sum\limits_{j = 1}^M { w_{ji}^2 {\left\| {{\mathbf{x}}_i} - {{\mathbf{d}}_j} \right\|}_2^2} + \mu {\left\| {\mathbf{W}} \right\|_{2,1}} , \\
    &  \rm{s.t.} \left\{ {\begin{array}{*{20}{c}} {{\mathbf{e}}^T {{\mathbf{w}}_i} = 1 \ \ i = 1, \ldots, N}\\ 
    {{w_{ji}} = 0\ \ \  {\rm{\ if\ }}{j} \notin \mathcal{N}_ {{\mathbf{x}}_i} } \ \ \   \end{array}} \right.  .
    \end{aligned}
\end{equation}
By solving this optimization problem, we can select the most representative landmarks to linearly approximate the entire dataset with minimal error.

Reconstructing the original CSI from the low-dimensional embedding is an inverse problem of dimensionality reduction. It can be described as how to discover the reconstruction function $g$ from the training data sets ${\mathbf{Y}}$ and ${\mathbf{X}}$. Likewise, let ${{\mathbf{D}}_{\rm{L}}^{\rm{rc}}} = \left[ {\begin{array}{*{20}{c}} {{{\mathbf{b}}}_1}&{{{\mathbf{b}}}_2}& \ldots &{{{\mathbf{b}}}_M} \end{array}} \right] \in {\mathbb{R}^{d \times M}}$ be a reconstruction dictionary in the low-dimensional space. And the corresponding dictionary in the high-dimensional space is ${\mathbf{D}}_{\rm{H}}^{\rm{rc}} = $ $ \left[ {\begin{array}{*{20}{c}} {g({{\mathbf{b}}_1})}&{g({{\mathbf{b}}_2})}& \ldots &{g({{\mathbf{b}}_M})} \end{array}} \right] \in {\mathbb{R}^{2{N_f} \times M}}$. In accordance with the idea of minimizing the local approximation error in Eq.~\eqref{dr_objective function}, the optimization problem of reconstruction can be formulated as
\begin{equation}
    \begin{aligned}
    \langle {\mathbf{D}}_{\rm{L}}^{\rm{rc}},{\mathbf{W}} \rangle = & \mathop {\arg \min }\limits_{{\mathbf{D}}_{\rm{L}}^{{\rm{rc}}},{\mathbf{W}}} {\left\| {\mathbf{Y}} - {\mathbf{D}}_{\rm{L}}^{\rm{rc}} {\mathbf{W}} \right\|_F^2} \\ 
    & + \lambda \sum\limits_{i = 1}^N {\left\|  {{\mathbf{x}}_i} - \sum\limits_{j = 1}^M {w_{ji} g( {{\mathbf{b}}_j} )} \right\|_2^2}  + \mu {\left\| {\mathbf{W}} \right\|_{2,1}}
    \end{aligned} ,
    \label{recon_objective function}
\end{equation}
which is similar to Eq.~\eqref{dr_objective function}, except that the parameters ${{\mathbf{x}}_i}$, ${{\mathbf{d}}_j}$ and the function $f$ are replaced by ${{\mathbf{y}}_i}$, ${{\mathbf{b}}_j}$ and $g$, respectively. In fact, it has been shown in part of our previous work \cite{cao2022manifold} that the solution to the reconstruction problem can be inferred directly from the solution to the dimensionality reduction problem. Therefore, the detailed derivation of the reconstruction problem is omitted. The next subsection only provides how to optimize the problem of dimensionality reduction.

\subsection{Proposed alternating iterative optimization algorithm}
Even though the optimization problem Eq.~\eqref{final objective function} is not jointly convex to $({{\mathbf{D}}_{\rm{H}}^{\rm{dr}}},{\mathbf{W}})$, it is convex to ${\mathbf{D}}_{\rm{H}}^{\rm{dr}}$ or ${\mathbf{W}}$ while the other is fixed. Therefore, we propose an alternating iteration optimization algorithm to split the joint optimization problem into two sub-problems. That is, the dictionary ${\mathbf{D}}_{\rm{H}}^{\rm{dr}}$ is fixed so that there is only one variable in Eq.~\eqref{final objective function}, making it convenient to calculate the optimal weight matrix ${\mathbf{W}}$. In a similar way, the dictionary is optimized by keeping ${\mathbf{W}}$ fixed. In addition, each sub-problem has a closed-form solution. In what follows, we provide the implementation details of the optimization algorithm. 

To begin with, suppose the weight matrix ${\mathbf{W}}$ has been initialized or updated in the previous iteration. We decompose the minimization problem of ${\mathbf{D}}_{\rm{H}}^{\rm{dr}}$ into sequential minimization problems \cite{sahoo2013dictionary}. Each column ${{\mathbf{d}}_j}$ in ${\mathbf{D}}_{\rm{H}}^{\rm{dr}}$ is updated separately to simplify the solution procedure. In light of this, the objective function with respect to ${{\mathbf{d}}_j}$ can be rewritten as
\begin{equation}
    \begin{aligned}
    \langle {{\mathbf{d}}_j} \rangle =  \mathop {\arg \min }\limits_{{\mathbf{d}}_j} {\left\| {\mathbf{E}}  - {{\mathbf{d}}_j}{{\mathbf{w}}_{j*}} \right\|_F^2} + \lambda \left\{ \sum\limits_{i = 1}^N {w_{ji}^2 {\left\| {{\mathbf{x}}_i} - {{\mathbf{d}}_j} \right\|^2}} \right. & \\ 
    \left. + \sum\limits_{i = 1}^N {\sum\limits_{m \ne j}^M {w_{mi}^2 {\left\| {{\mathbf{x}}_i} - {{\mathbf{d}}_m} \right\|^2 } }}  \right\} + \mu {\left\| {\mathbf{W}} \right\|_{2,1}} & 
    \end{aligned},
    \label{dr_high dictionary objective function}
\end{equation}
where  ${\mathbf{E}} = {\mathbf{X}} - \sum\limits_{m \ne j}^M {{{\mathbf{d}}_m}{{\mathbf{w}}_{m*}}}$, and ${{\mathbf{w}}_{j*}} \in {\mathbb{R}^{1 \times {N}}}$ is the $j$-th row of ${\mathbf{W}}$.

\begin{theorem}
\label{thm1}
The optimal solution to the problem Eq.~\eqref{dr_high dictionary objective function} is
\begin{equation}
    {{\mathbf{d}}_j} = \frac{{\mathbf{E}} {\mathbf{w}}_{j*}^T + \lambda {\mathbf{X}}{{({\mathbf{w}_{j*}^ 2} )}^T}} {( {1 + \lambda}) {{\mathbf{w}_{j*}} {\mathbf{w}}_{j*}^T} },
    \label{dr_solution to high dictionary}
\end{equation}
where ${\mathbf{w}}_{j*}^2$ represents the square of each entry in ${\mathbf{w}}_{j*}$.

\begin{prof}
: \rm{Please refer to Appendix \ref{appendixB}.}
\end{prof}
\end{theorem}

In the weight update stage, fix the dictionary ${\mathbf{D}}_{\rm{H}}^{\rm{dr}}$ after it has been initialized or updated. Based on the $k$-NN criterion, the $k$ nearest landmarks of ${{\mathbf{x}}_i}$ can be determined in terms of Euclidean distance and form a neighborhood matrix ${\mathbf{N}}_ {{\mathbf{x}}_i} = \left[ {\begin{array}{*{20}{c}} {{\mathbf{d}}_{a_1}}&{{\mathbf{d}}_{a_2}}& \ldots &{{\mathbf{d}}_{a_k}} \end{array}} \right] \in {\mathbb{R}^{2{N_f} \times k}}$. The corresponding column indices make up an index vector ${\mathbf{a}} = $ $\left[ {\begin{array}{*{20}{c}} {a_1}&{a_2}& \ldots &{a_k} \end{array}} \right] \in {\mathbb{R}^{1 \times k}}$. In fact, the weight vector ${{\mathbf{w}}_i}$ is sparse and contains only $k$ non-zero entries. Let ${\hat {\mathbf{w}}_i} ={\left[ {\begin{array}{*{20}{c}} { \hat w_{1i}}&{\hat w_{2i}}& \ldots &{\hat w_{ki}} \end{array}} \right]^T} $ $\in {\mathbb{R}^{k \times 1}}$ be a compact sub-vector of non-zero entries in ${{\mathbf{w}}_i}$, where ${\hat w}_{ji}$ is equal to ${w}_{(a_j)i}$, $j=1, \ldots, k$. By replacing ${{\mathbf{w}}_i}$ with ${\hat {\mathbf{w}}_i}$ and dropping ${\mathbf{D}}_{\rm{H}}^{\rm{dr}}$, the optimization problem about ${\hat {\mathbf{w}}_i}$ is simplified to
\begin{equation}
    \begin{aligned}
    \langle {\mathbf{W}} \rangle = & \mathop {\arg \min}\limits_{\mathbf{W}} \sum\limits_{i = 1}^N \left({\left\| {{{\mathbf{x}}_i} - {\mathbf{N}}_{{\mathbf{x}}_i}{{\hat {\mathbf{w}}}_i}} \right\|_2^2} \right. \\
    & \ \ \ \ \ \ \ \ \ \ \ \ \ \ \ \  + \lambda \sum\limits_{j = 1}^k {{\hat w}_{ji}^2} {\left. {\left\| {{\mathbf{x}}_i} - {{\mathbf{d}}_{{a_j}}} \right\|_2^2} \right)} + \mu {\left\| {\mathbf{W}} \right\|_{2,1}}  
    \end{aligned} ,
    \label{dr_coding objective function}
\end{equation}
where the weights still satisfy the sum-to-one constraint.

In general, it is challenging to minimize the $L_{2,1}$-norm. Fortunately, the derivative of ${\left\| {\mathbf{W}} \right\|_{2,1}}$ can be easily accessible. Its equivalent is the derivative of ${\rm{tr}}({\mathbf{W}}^T{\mathbf{G}}{\mathbf{W}})$ \cite{nie2010efficient}, where ${\mathbf{G}}$ is a diagonal matrix with the $i$-th diagonal element being
\begin{equation}
    \label{g}
    g_{ii} = \frac{1}{2{\| {{\mathbf{w}}_{i*}} \|_2}}.
\end{equation}

\begin{theorem}
\label{thm2}
The optimal ${\hat {\mathbf{w}}_i}$ to the problem Eq.~\eqref{dr_coding objective function} is
\begin{equation}
    {\hat {\mathbf{w}}_i} =  \frac{{{{\left( {\mathbf{R}} + \lambda \varphi({\mathbf{R}}) + \mu{\hat{ \mathbf{G}}} \right)}^{ - 1}}{\mathbf{e}}}}{{{{\mathbf{e}}^T}{{\left( \mathbf{R} + \lambda \varphi({\mathbf{R}}) + \mu{\hat{ \mathbf{G}}} \right)}^{ - 1}}{\mathbf{e}}}}
    \label{dr_solution to coding},
\end{equation}
where ${\mathbf{R}} = {({{\mathbf{x}}_i}{{\mathbf{e}}^T} - {\mathbf{N}}_{{\mathbf{x}}_i})^T} ( {{\mathbf{x}}_i}{{\mathbf{e}}^T} -{\mathbf{N}}_{{\mathbf{x}}_i})$, $\varphi ({\mathbf{R}}) $ is a matrix that only preserves the diagonal elements of the matrix ${\mathbf{R}}$ and sets the rest to zero, and ${\hat{ \mathbf{G}}} = {\rm{diag}}(g_{a_1 a_1},g_{a_2 a_2},\ldots,g_{a_k a_k})$.

\begin{prof}
: \rm{Please refer to Appendix \ref{appendixC}.}
\end{prof}
\end{theorem}

According to the above closed-form solutions, ${\mathbf{W}}$ and ${\mathbf{D}}_{\rm{H}}^{\rm{dr}}$ can be optimized alternatively. Once Eq.~\eqref{final objective function} falls below a predetermined threshold or the maximum number of iterations is reached, the iteration process will be terminated. Afterwards, the low-dimensional dictionary ${\mathbf{D}}_{\rm{L}}^{\rm{dr}}$ is obtained by solving
\begin{equation}
    \mathop {\min }\limits_{{\mathbf{D}}_{\rm{L}}^{\rm{dr}}} \sum\limits_{i = 1}^{N} {\left\| {\mathbf{y}_i}  - \sum\limits_{j = 1}^M {w_{ji}}f\left( {{\mathbf{d}}_j} \right)  \right\|_2^2} =  {\left\| {{\mathbf{Y}} - {\mathbf{D}}_{\rm{L}}^{\rm{dr}}{\mathbf{W}}} \right\|_F^2} ,
    \label{dr_low dictionary objective function}
\end{equation}
which is based on the fact that the local geometric relationships between ${\mathbf{x}_i}$ and ${\mathbf{D}}_{\rm{H}}^{\rm{dr}}$ on $\mathcal{M}$ as well as those between ${\mathbf{y}_i}$ and ${\mathbf{D}}_{\rm{L}}^{\rm{dr}}$ on $\mathcal{S}$ are characterized by the same set of local weights. The least squares solution to Eq.~\eqref{dr_low dictionary objective function} is
\begin{equation}
   {\mathbf{D}}_{\rm{L}}^{\rm{dr}} = {\mathbf{Y}}{{\mathbf{W}}^T}{( {\mathbf{W}{{\mathbf{W}}^T}})^{-1}}.
    \label{dr_solution to low dictionary}
\end{equation}

Until now, both the high-dimensional dictionary ${\mathbf{D}}_{\rm{H}}^{\rm{dr}}$ and the low-dimensional dictionary ${\mathbf{D}}_{\rm{L}}^{\rm{dr}}$ are available for dimensionality reduction, which are pre-stored at the UE side to calculate the embedding of the incremental CSI. The landmark selection method is summarized in Algorithm~\ref{alg1}. Note that the weight matrix is an intermediate variable that might be discarded. Similarly, the group of reconstruction dictionaries, namely ${{\mathbf{D}}_{\rm{H}}^{\rm{rc}}}$ and ${{\mathbf{D}}_{\rm{L}}^{\rm{rc}}}$, can also be obtained, enabling the BS to recover the high-dimensional CSI from the low-dimensional embedding.

\subsection{Convergence analysis for the proposed algorithm}
In this section, we will analyze the convergence of the proposed Algorithm~\ref{alg1}. Here, Lemma \ref{lemma2} \cite{nie2010efficient} is introduced to assist the analysis.

\begin{lemma}
    \label{lemma2}
    For any non-zero vectors ${\mathbf{s}}$, ${\mathbf{v}} \in {\mathbb{R}^n}$, the following inequality holds 
    \begin{eqnarray*}
        \frac{\left\| {\mathbf{s}} \right\|_2^2}{2{\left\| {\mathbf{s}} \right\|}_2} - {\left\| {\mathbf{s}} \right\|_2} \le \frac{\left\| {\mathbf{v}} \right\|_2^2}{2{\left\| {\mathbf{s}} \right\|}_2} - {\left\| {\mathbf{v}} \right\|_2} .
    \end{eqnarray*}
\end{lemma}

The convergence of Algorithm~\ref{alg1} is summarized in the following theorem. To simplify the notation, both the superscript $(\cdot)^{\rm{dr}}$ and the subscript $(\cdot)_{\rm{H}}$ of ${\mathbf{D}}_{\rm{H}}^{\rm{dr}}$ are dropped. 

\begin{theorem}
\label{thm3}
In each iteration, the value of the objective function Eq.~\eqref{final objective function} decreases monotonically:
\begin{equation*}
    {U({\mathbf{D}_t}, {\mathbf{W}_t})} + \mu {\left\| \mathbf{W}_t \right\|_{2,1}} \ge {U({\mathbf{D}_{t+1}}, {\mathbf{W}_{t+1}})} + \mu {\left\| \mathbf{W}_{t+1} \right\|_{2,1}},
\end{equation*}
where in the $t$-th iteration, ${U({\mathbf{D}_t}, {\mathbf{W}_t})} = \left\| {\mathbf{X} - {\mathbf{D}_t} {\mathbf{W}_t}} \right\|_F^2 + \lambda \sum\limits_{i = 1}^N \sum\limits_{j = 1}^M (w_{ji}^T)^2 {\left\| {{\mathbf{x}}_i} - {\mathbf{d}}_j^T \right\|}_2^2.$
\end{theorem}

\begin{prof}
: Since the derivative of ${\left\| {\mathbf{W}} \right\|_{2,1}}$ equals the derivative of ${\rm{tr}}({\mathbf{W}}^T{\mathbf{G}}{\mathbf{W}})$, it can be easily verified that the solution to \eqref{final objective function} is the solution to the following problem
\begin{equation}
     \mathop {\min} \limits_{{\mathbf{D}},{\mathbf{W}}}  {U({\mathbf{D}}, {\mathbf{W}})} + \mu {\rm{tr}}({\mathbf{W}}^T{\mathbf{G}}{\mathbf{W}}).
\end{equation}
Thus in the $t$-th iteration,
\begin{equation}
    \left\{ {\begin{array}{*{20}{l}}
    {\mathbf{D}_{t+1} = \mathop {\arg \min} \limits_{\mathbf{D}}  U({\mathbf{D}}, {\mathbf{W}_t}) + \mu {\rm{tr}}({\mathbf{W}_t^T}{\mathbf{G}_t}{\mathbf{W}_t})} \\
    {\mathbf{W}_{t+1} = \mathop {\arg \min} \limits_{\mathbf{W}}  {U({\mathbf{D}_{t+1}}, {\mathbf{W}})} + \mu {\rm{tr}}({\mathbf{W}}^T{\mathbf{G}_t}{\mathbf{W}}) }
\end{array}} \right. .
\end{equation}

According to the convex optimization theory \cite{boyd2004convex}, we have
\begin{equation}
\label{inequality}
\begin{aligned}
    &{U({\mathbf{D}_t}, {\mathbf{W}_t})}
    + \mu {\rm{tr}}(\mathbf{W}_t^T{\mathbf{G}_t}{\mathbf{W}_t}) \\ & \ge {U({\mathbf{D}_{t+1}}, {\mathbf{W}_{t+1}})} 
    + \mu {\rm{tr}}(\mathbf{W}_{t+1}^T{\mathbf{G}_t}{\mathbf{W}_{t+1}}) 
\end{aligned}.
\end{equation}
On the other hand, the right side of the above inequality meets
\begin{equation}
\label{left hand}
\begin{aligned}     
    &{U\left({\mathbf{D}_{t+1}}, {\mathbf{W}_{t+1}} \right)} + \mu {\rm{tr}}(\mathbf{W}_{t+1}^T{\mathbf{G}_t}{\mathbf{W}_{t+1}})  \\
    & \mathop = \limits^{(a)} {U \left({\mathbf{D}_{t+1}}, {\mathbf{W}_{t+1}}\right)} + \mu {\rm{tr}}(\mathbf{W}_{t+1}^T{\mathbf{G}_{t+1}}{\mathbf{W}_{t+1}}) \\ 
    & \ \ \ + \mu \sum\limits_{i = 1}^M {\left(\frac{{\left\| {\mathbf{W}_{i*}^{t + 1}} \right\|}_2^2}{2\left\| {\mathbf{W}_{i*}^T} \right\|_2} - \frac{1}{2}\left\| {\mathbf{W}_{i*}^{t+1}} \right\|_2 \right)} \\  
    & \mathop \ge \limits^{(b)} 
    {U \left({\mathbf{D}_{t+1}}, {\mathbf{W}_{t+1}}\right)} + \mu {\rm{tr}}(\mathbf{W}_{t+1}^T{\mathbf{G}_{t+1}}{\mathbf{W}_{t+1}}) \\  
    & \ \ \  + \mu \sum\limits_{i = 1}^M {\left( \frac{1}{2} \left\| {\mathbf{W}_{i*}^{t + 1}} \right\|_2 - \frac{1}{2}{\left\| {\mathbf{W}_{i*}^T} \right\|_2}\right)} 
    \end{aligned} \ ,
\end{equation}
where, according to the definition of the matrix trace, we have ${\rm{tr}}({\mathbf{W}^T}\mathbf{G} \mathbf{W}) = \sum\limits_{i = 1}^M {{g_{ii}}{\left\| {\mathbf{W}_{i*}} \right\|}_2^2}$ in $(a)$, and $(b)$ is derived from Lemma \ref{lemma2}. 

Combining Eq.~\eqref{inequality} and Eq.~\eqref{left hand}, we have
\begin{equation}
    \begin{aligned}     
    & {U({\mathbf{D}_t}, {\mathbf{W}_t})} 
    + \mu {\rm{tr}}(\mathbf{W}_t^T{\mathbf{G}_t}{\mathbf{W}_t}) \\ 
    & \ge {U({\mathbf{D}_{t+1}}, {\mathbf{W}_{t+1}})} + \mu {\rm{tr}}(\mathbf{W}_{t+1}^T{\mathbf{G}_{t+1}}{\mathbf{W}_{t+1}}) \\ 
    & \ \ \ + \mu \sum\limits_{i = 1}^M {\left( \frac{1}{2} \left\| {\mathbf{W}_{i*}^{t + 1}}  \right\|_2 - \frac{1}{2}{\left\| {\mathbf{W}_{i*}^T} \right\|_2} \right)} \\ 
    & \mathop = \limits^{(c)} {U({\mathbf{D}_{t+1}}, {\mathbf{W}_{t+1}})} + 2\mu {\rm{tr}}(\mathbf{W}_{t+1}^T{\mathbf{G}_{t+1}}{\mathbf{W}_{t+1}}) \\ 
    & \ \ \ - {\frac{\mu}{2}} \sum\limits_{i = 1}^M {\left\| {\mathbf{W}_{i*}^T} \right\|_2} &
    \end{aligned} ,
    \label{shift inequality}
\end{equation}
where $(c)$ uses the definition of $\mathbf{G}$ in Eq.~\eqref{g}. Further, the following inequality holds 
\begin{equation}
\begin{aligned}     
    &{U({\mathbf{D}_t}, {\mathbf{W}_t})} 
    + 2 \mu {\rm{tr}}(\mathbf{W}_t^T{\mathbf{G}_t}{\mathbf{W}_t}) 
    \\ & = {U({\mathbf{D}_t}, {\mathbf{W}_t})} 
    + \mu {\rm{tr}}(\mathbf{W}_t^T{\mathbf{G}_t}{\mathbf{W}_t}) + {\frac{\mu}{2}} \sum\limits_{i = 1}^M {\left\| {\mathbf{W}_{i*}^T} \right\|_2} 
    \\ & \ge  
    {U({\mathbf{D}_{t+1}}, {\mathbf{W}_{t+1}})} + 2\mu {\rm{tr}}(\mathbf{W}_{t+1}^T{\mathbf{G}_{t+1}}{\mathbf{W}_{t+1}}) 
    \end{aligned} .
\end{equation}
That is to say, 
\begin{equation}
    {U({\mathbf{D}_t}, {\mathbf{W}_t})} + \mu {\left\| \mathbf{W}_t \right\|_{2,1}} \ge {U({\mathbf{D}_{t+1}}, {\mathbf{W}_{t+1}})} + \mu {\left\| \mathbf{W}_{t+1} \right\|_{2,1}},
\end{equation}
which indicates that the value of the objective function Eq.~\eqref{final objective function} will decrease monotonically in each iteration. 
\end{prof}

\begin{algorithm}[t]
\caption{Landmark selection for dimensionality reduction} 
\label{alg1}
\hspace*{0.02in} {\bf Input:} 
The historical CSI data set ${\mathbf{X}}$, the size of dictionary $M$, the number of nearest landmarks $k$, the intrinsic dimensionaity $d$, $\lambda$, $\mu$.\\
\hspace*{0.02in} {\bf Output:} 
A group of dictionaries ${\mathbf{D}}_{\rm{H}}^{\rm{dr}}$ and ${\mathbf{D}}_{\rm{L}}^{\rm{dr}}$.
\begin{algorithmic}[1]
\State $t=0$, initialize $({\mathbf{D}}_{\rm{H}}^{\rm{dr}})_t$ by randomly selecting $M$ columns from ${\mathbf{X}}$, initialize ${\mathbf{W}_t}$;
\While{Eq.~\eqref{final objective function} converges}
    \State Update ${\mathbf{G}_t}$ based on Eq.~\eqref{g};
    \For{each landmark $\mathbf{d}_j$, $j = 1, \ldots ,M$}  \State Compute the $j$-th column of $({\mathbf{D}}_{\rm{H}}^{\rm{dr}})_{t+1}$ by Eq.~\eqref{dr_solution to high dictionary};
    \EndFor
    \For{each sample ${\mathbf{x}}_i$, $i = 1, \ldots ,N$} \State Determine the neighborhood matrix ${\mathbf{N}}_ {{\mathbf{x}}_i} $;
    \State Compute the weight vector ${{\mathbf{w}}_i^{t+1}}$ by Eq.~\eqref{dr_solution to coding};
    \EndFor
    \State $t \leftarrow t+1$;
\EndWhile
\State Compute the low-dimensional embedding ${\mathbf{Y}}$ using a ML algorithm;
\State Compute the low-dimensional dictionary ${\mathbf{D}}_{\rm{L}}^{\rm{dr}}$ by Eq.~\eqref{dr_solution to low dictionary};
\State \Return ${\mathbf{D}}_{\rm{H}}^{\rm{dr}}$, ${\mathbf{D}}_{\rm{L}}^{\rm{dr}}$
\end{algorithmic}
\end{algorithm}

\section{MLCF-based CSI Feedback}
In previous sections, we have obtained two sets of dictionaries, one for compression and the other for reconstruction. It is worth mentioning that these two sets of dictionaries are learned at the BS beforehand from the training data sets $\mathbf{X}$ and $\mathbf{Y}$. Once both sets are known, the BS will store the group of reconstruction dictionaries itself and broadcast this group of compression dictionaries to the UE. Below, we will show how to compress the newly sampled CSI at the UE and reconstruct the original CSI at the BS with these dictionaries. 

\subsection{Dimensionality reduction for the incremental CSI}
In FDD massive MIMO systems, we hope to reduce the downlink CSI feedback overhead at the UE while accurately reconstructing the CSI at the BS. With that in mind, the UE adopts the group of aforementioned compression dictionaries, namely ${\mathbf{D}}_{\rm{H}}^{\rm{dr}}$ and ${\mathbf{D}}_{\rm{L}}^{\rm{dr}}$, to compute the low-dimensional embedding of the downlink CSI. 

Let the incremental CSI, estimated from the downlink pilots at a new time slot, be denoted by
\begin{align*} 
{ \widetilde {\mathbf{H}}_{\rm{new}} = \left[ {\begin{array}{*{20}{c}} {\mathbf{h}}_1& \ldots & {\mathbf{h}}_{N_t} \end{array}} \right] \in {\mathbb{R}^{2{N_f} \times {N_t}}} }.
\end{align*}
Since the landmarks on the manifold $\mathcal{M}$ have been learned, the local geometric property of $\widetilde {\mathbf{H}}_{\rm{new}}$ on $\mathcal{M}$ can be easily characterized by its nearest neighbors in the landmarks. 

In the process of dimensionality reduction, we first search for the $k$ nearest neighbors of ${\mathbf{h}}_i$, $i =1,2,\ldots,{N_t}$, in ${\mathbf{D}}_{\rm{H}}^{\rm{dr}}$ and model the local geometries on $\mathcal{M}$ as a collection of linear patches. The geometries are expected to remain consistent in both high-dimensional and low-dimensional spaces. Thus, the weight and neighborhood relationships can be settled by optimizing the following objective function
\begin{equation}
    \label{increment_dr objective function}
    \begin{aligned}
    & \mathop {\min }\limits_{{\mathbf{w}}_i} {\left\| {\mathbf{h}}_i - {\mathbf{D}}_{\rm{H}}^{\rm{dr}} {{\mathbf{w}}_i} \right\|^2} + \lambda {\left\| f({\mathbf{h}}_i) - \sum\limits_{j = 1}^M {w_{ji}} f({{\mathbf{d}}_j})  \right\|^2} \\
    & \rm{\ s.t.} \left\{ {\begin{array}{*{20}{c}} {{\mathbf{e}}^T {{\mathbf{w}}_i} = 1\ \  i = 1, \ldots ,N_t}\\ 
    { w_{ji} = 0\ \ \ \ {\rm{\ if\ }}{j} \notin \mathcal{N}_ {{\mathbf{h}}_i} } \ \ \ \ \end{array}} \right.  ,
    \end{aligned}
\end{equation}
which is analogous to Eq.~\eqref{dr_objective function}, except that the dictionary ${\mathbf{D}}_{\rm{H}}^{\rm{dr}}$ is already known and $\mu$ is a special case equal to 0. Referring to the derivation in Appendix \ref{appendixC}, the optimal solution to the $k$ non-zero entries in ${\mathbf{w}}_i$ is
\begin{equation}
    {\hat {\mathbf{w}}_i} = \frac{{{ \left( {{\mathbf{R}} + \lambda \varphi ({\mathbf{R}}) } \right)}^{-1}}{\mathbf{e}}}{{{\mathbf{e}}^T}{{\left( {\mathbf{R}} + \lambda \varphi ({\mathbf{R}})  \right)}^{ - 1}}{\mathbf{e}}}
    \label{increment_solution to coding},
\end{equation}
where \[{\mathbf{R}} ={({{\mathbf{h}}_i}{{\mathbf{e}}^T} - \mathbf{N}_{{\mathbf{h}}_i})^T} ({{\mathbf{h}}_i}{{\mathbf{e}}^T} - {\mathbf{N}}_{{\mathbf{h}}_i}).\] With $i$ ascending from 1 to $N_t$, we obtain the whole weight matrix ${\mathbf{W}}_{\rm{dr}} = \left[ {\begin{array}{*{20}{c}} {{\mathbf{w}}_1}& {{\mathbf{w}}_2}& \ldots &{{\mathbf{w}}_{N_t}} \end{array}} \right] \in {\mathbb{R}^{M \times {N_t}}}$. Since the weights ${\mathbf{W}}_{\rm{dr}}$ can also characterize the local geometric property in the low-dimensional space, the low-dimensional embedding of $\widetilde {\mathbf{H}}_{\rm{new}}$ can be calculated by 
\begin{equation}
    {{\mathbf{Y}}_{\rm{new}}} = {\mathbf{D}}_{\rm{L}}^{\rm{dr}}{{\mathbf{W}}_{\rm{dr}}}
    \label{increment_low embedding}.
\end{equation}
Then, the UE feeds the low-dimensional embedding ${\mathbf{Y}}_{\rm{new}} = \left[ {\begin{array}{*{20}{c}} {{\mathbf{y}}_1}& {{\mathbf{y}}_2}& \ldots &{{\mathbf{y}}_{N_t}} \end{array}} \right]\in {\mathbb{R}^{d \times {N_t}}}$ back to the BS without additional parameters, which is convenient in practical applications.

Compressing the incremental CSI with landmarks has the advantage of low computational complexity. On the one hand, when the new CSI arrives, only the low-dimensional embedding of $\widetilde {\mathbf{H}}_{\rm{new}}$ is calculated. There is no need to rerun the entire ML algorithms with the original data set augmented by the new sample. On the other hand, compared to the conventional ML algorithms that operate on the entire data set, it is easier to obtain the neighborhood and weight relationships between the CSI and the landmarks. The time complexity for computing a weight vector is $O(2N_f k^2)+O(k^3)$, which is dominated by $O(2N_f k^2)$ as $k \ll 2N_f$. In additional, computing ${\mathbf{Y}}_{\rm{new}}$ has a time complexity of $O(dMN_t)$. Hence, the total complexity of dimensionality reduction is $O(2N_f N_t k^2)+ O(dMN_t)$. 

\subsection{Reconstruction for the incremental CSI}
After receiving the low-dimensional embedding ${\mathbf{Y}}_{\rm{new}}$, the BS attempts to reconstruct the CSI ${\hat {\mathbf{H}}_{\rm{new}}}$ as close as possible to the true value ${ {\mathbf{H}}_{\rm{new}}}$. Perfectly reconstructing the CSI is not feasible since some information is lost during the process of dimensionality reduction. In order to simplify the reconstruction while guaranteeing the reconstruction quality, we provide a low-complexity reconstruction scheme based on the selected landmarks. Inspired by the process of compressing the incremental CSI, we believe that the reconstruction mapping may still be established by keeping the local geometry on the manifold unchanged. For the newly received ${\mathbf{Y}}_{\rm{new}}$, its local geometric relationship with ${\mathbf{D}}_{\rm{L}}^{\rm{rc}}$ can be easily identified. Under this premise, the reconstructed ${\hat {\mathbf{H}}_{\rm{new}}}$ can be obtained by maintaining the same geometric relationship with ${\mathbf{D}}_{\rm{H}}^{\rm{rc}}$ in the high-dimensional space.

Firstly, search for the $k$ nearest neighbors of ${{\mathbf{y}}_i}$ in the pre-stored dictionary ${\mathbf{D}}_{\rm{L}}^{\rm{rc}}$. Based on the idea that keeps the local geometric property constant, the optimization problem with respect to the reconstruction weights can be formulated as
\begin{equation}
    \begin{aligned}
    & \mathop {\min }\limits_{{\mathbf{w}}_i} {\left\| {{{\mathbf{y}}_i} - {\mathbf{D}}_{\rm{L}}^{{\rm{rc}}}{{\mathbf{w}}_i}}  \right\|^2} + \lambda {\left\| {g({{\mathbf{y}}_i} ) - \sum\limits_{j = 1}^M {{{w}_{ji}}g({\mathbf{b}}_j)} } \right\|^2} \\
    & \rm{\ s.t.} \left\{ {\begin{array}{*{20}{c}} {{\mathbf{e}}^T {{\mathbf{w}}_i} = 1\ \  i = 1, \ldots ,N_t}\\
    { w_{ji} = 0\ \ \ \ {\rm{\ if\ }}{j} \notin \mathcal{N}_ {{\mathbf{y}}_i} } \ \ \ \ \end{array}} \right. ,
    \end{aligned}
    \label{increment_recon objective function}
\end{equation}
where ${\mathbf{w}}_i$ is the $i$-th column of the reconstruction weight matrix ${\mathbf{W}}_{\rm{rc}} = \left[ {\begin{array}{*{20}{c}} {{\mathbf{w}}_1}& {{\mathbf{w}}_2}& \ldots &{{\mathbf{w}}_{N_t}} \end{array}} \right] \in {\mathbb{R}^{M \times {N_t}}}$. It can be observed that the objective function Eq.~\eqref{increment_recon objective function} is comparable to Eq.~\eqref{increment_solution to coding} except for some parameters. Hence, the $k$ non-zeros entries of  ${\mathbf{w}}_i$ can likewise be computed by Eq.~\eqref{increment_solution to coding}, where ${\mathbf{R}}$ is reset to $ {(  {{\mathbf{y}}_i}{{\mathbf{e}}^T} - {\mathbf{N}}_{{\mathbf{y}}_i})^T} ( {{\mathbf{y}}_i}{{\mathbf{e}}^T} - {\mathbf{N}}_{{\mathbf{y}}_i})$. 
Subsequently, based on ${\mathbf{W}}_{\rm{rc}}$ and the known dictionary ${\mathbf{D}}_{\rm{H}}^{\rm{rc}}$, the BS reconstructs ${\hat {\mathbf{H}}_{\rm{new}}}$ by
\begin{equation}
    {\hat {\mathbf{H}}_{\rm{new}}} = {{\mathbf{D}}_{\rm{H}}^{\rm{rc}}}{{\mathbf{W}}_{\rm{rc}}}.
    \label{increment_reconstruction}
\end{equation}

Utilizing the knowledge of manifold structure makes CSI reconstruction simple. Moreover, the above process only requires vector and matrix calculations, rather than multiple iterations. The overall time complexity of reconstruction is $O(N_t k^3) + O(dN_t k^2)+ O(2N_fMN_t)$. At this point, the tasks of compressing and reconstructing the incremental CSI have been accomplished.

\section{Numerical Results}\label{Sec:Sim}
In this section, we evaluate the performance of the proposed MLCF method under the industrial CDL channel model, conforming to 3GPP TR 38.901 \cite{3GPP2019Study}. With the help of MATLAB, we carry on the simulation for the 5G new radio (NR) Release 16 and generate the training and testing data sets \cite{wang2021compressive}. %\footnote{We generate the data sets following the open source code on Github: https://github.com/CodeDwan/EMEV-feedback \cite{zhang2022attention}. }. 
Table.~\ref{tab1} shows part of the default parameters in the channel model. For an OFDM system, it is necessary to consider multiple subcarriers and OFDM symbols. In practice, $N_f$ can also denote the number of resource blocks (RBs) or the number of groups of consecutive RBs. In this paper, RB is adopted as the basic feedback granularity. Unless particularly specified, the constant $\lambda$, $\mu$, the number of nearest neighbors $k$, the size of dictionary $M$, the size of data set $N$, and the number of BS antenna $N_t$ are set to 0.001, 0.001, 70, 400, 4000, and 32, respectively. The $k$-nearest neighbor ($k$-NN) strategy is adopted to select the neighbors in landmarks.

\begin{table}[htbp]
\setlength{\abovecaptionskip}{0pt}    
\setlength{\belowcaptionskip}{10pt}
\caption{The parameter settings for CDL channel}
\label{tab1}
\setlength{\tabcolsep}{4mm}
\begin{center}
    \begin{tabular}{|c|c|}
    \hline
    Parameters  & Default values \\ \hline
    Channel model  &   CDL-A     \\ \hline
    Number of clusters   & $N_c$ = 23    \\ \hline
    Total number of rays  & 460    \\ \hline
    Delay spread   & 300 ns     \\ \hline
    Time slot  &   1 ms    \\ \hline
    Downlink carrier frequency  &  3.5 GHz     \\ \hline
    Subcarrier spacing  & $\Delta f$ = 15 kHz     \\ \hline
    Resource blocks  &  $N_{RB}$ = 48     \\ \hline
    UE speed   & 30 km/h    \\ \hline
    Number of BS antennas  &   $N_t$ = 32/64    \\ \hline
    \end{tabular}
\end{center}
\end{table}

To analyze the reconstruction performance, we introduce two evaluation metrics. The first metric NMSE measures the difference between the reconstructed CSI and the original CSI, defined as 
\begin{equation}
    {\rm{NMSE}} = 10 \lg \left\{ \mathbb{E} \frac{{\| {\widetilde{\mathbf{H}}} - {\hat {\mathbf{H}}} \|_F^2}}
    {\| {\widetilde {\mathbf{H}}} \|_F^2} \right\}.
    \label{NMSE}
\end{equation}
The other one is cosine similarity, defined as
\begin{equation}
   \rho = \mathbb{E} \left\{ {\frac{1}{N_f}\sum\limits_{n = 1}^{N_f} {\frac{| {\hat {\mathbf{h}}}{_n} {{\mathbf{h}}_n^H} |}  {{\| {\hat {\mathbf{h}}}_n \|}_2 {\| {\mathbf{h}}_n \|}_2}} } \right\},
    \label{cosine similarity}
\end{equation}
where ${{\mathbf{h}}_n}$ and ${{\hat {\mathbf{h}}}_n}$ are the $n$-th subcarrier of the true complex channel and reconstructed channel, respectively.

Three DL-based algorithms (CLNet \cite{ji2021clnet}, NR-CsiNet \cite{zimaglia2020novel}, FISTA-Net \cite{guo2021csi}) are introduced as benchmarks, all of which employ a two-step compression operation. Specifically, the CSI matrix ${\mathbf{H}}\in {\mathbb{C}^{{N_f} \times {N_t}}}$ is transformed by discrete Fourier transformation (DFT) into the angular-delay domain $\mathbf{\bar{H}}$, which exhibits sparsity with only $N_a$ rows composed of non-zero values. First, ${\mathbf{\bar{H}}}$ is compressed to ${\mathbf{H}}_a \in {\mathbb{C}^{{N_a} \times {N_t}}} $ by retaining the $N_a$ non-zero rows of $\mathbf{\bar{H}}$ and removing the remaining rows. The compression ratio of this step is $\gamma_1 = {N_a \mathord{\left/ {\vphantom {N_a N_f}} \right. \kern-\nulldelimiterspace} N_f} $. Next, the benchmarks achieve the second step of compression by designing an encoder network to compress ${\mathbf{H}}_a$ into a codeword vector ${\mathbf{c}}\in {\mathbb{R}^{{L} \times {1}}}$. The compression ratio is $\gamma_2 = {L \mathord{\left/ {\vphantom {L 2N_a N_t}} \right. \kern-\nulldelimiterspace} 2N_a N_t} $. Meanwhile, a decoder network is deployed at the BS to recover ${\mathbf{\hat {H}}}_a$ from the codeword. Finally, the CSI matrix is recovered by performing inverse DFT. The total compression ratio is $\gamma = \gamma_1 \gamma_2 = {L \mathord{\left/ {\vphantom {L 2N_f N_t}} \right. \kern-\nulldelimiterspace} 2N_f N_t} $. To enable that our proposed MLCF has the same compression ratio as the benchmarks, we set $\gamma_1 = {1 \mathord{\left/ {\vphantom {1 2}} \right. \kern-\nulldelimiterspace} 2} $ and ${\gamma _2} = \left[ {{1 \mathord{\left/ {\vphantom {1 2}} \right. \kern-\nulldelimiterspace} 2},{1 \mathord{\left/ {\vphantom {1 4}} \right. \kern-\nulldelimiterspace} 4},{1 \mathord{\left/ {\vphantom {1 8}} \right. \kern-\nulldelimiterspace} 8}} \right]$. Note that the three network structures described in the related papers are maintained with the exception of the input and output dimensions. The training data set, validation data set, and testing data set contain 100,000, 30,000, and 20,000 samples respectively. For consistent comparisons, the testing set that MLCF and the benchmark algorithms employ is the same. 

\begin{table}[!ht]
\centering
\setlength{\abovecaptionskip}{0pt}    
\setlength{\belowcaptionskip}{10pt}
\caption{Performance (NMSE In $\rm{d}$B, Cosine Similarity) And Complexity Comparisons}
\label{tab2}
    \begin{tabular}{c|c|cc|cc}
    \hline
    \multirow{2}{*}[-1.5ex]{$\gamma$} & \multirow{2}{*}[-1.5ex]{Methods}  & \multicolumn{2}{c|}{Performance}  & \multicolumn{2}{c}{Complexity} \\ 
    \cline{3-6}  &  & \multicolumn{1}{c|}{NMSE} & $\rho$  & \multicolumn{1}{c|}{\begin{tabular}[c]{@{}c@{}} Trainable \\ Parameters \end{tabular}} & FLOPs  \\ 
    \hline
    1/16  & \begin{tabular}[c]{@{}c@{}} CLNet\\ NR-CsiNet \\ FISTA-Net\\ MLCF\end{tabular}
          & \multicolumn{1}{c|}{\begin{tabular}[c]{@{}c@{}} -16.76 \\ -19.97 \\ -22.88 \\ \textbf{-23.74} \end{tabular} }
          & \begin{tabular}[c]{@{}c@{}} 0.9905 \\ 0.9960 \\ 0.9981 \\ \textbf{0.9982} \end{tabular} 
          & \multicolumn{1}{c|}{ \begin{tabular}[c]{@{}c@{}} 593.798K \\ 593.336K \\ 331.672K \\ \textbf{81.6K}  \end{tabular}}
          & \begin{tabular}[c]{@{}c@{}} \textbf{2.332M} \\  3.279M \\ 26.428M \\ 23.706M
          \end{tabular} 
          \\ \hline
    1/8  & \begin{tabular}[c]{@{}c@{}} CLNet\\ NR-CsiNet \\ FISTA-Net\\ MLCF\end{tabular} 
          & \multicolumn{1}{c|}{\begin{tabular}[c]{@{}c@{}} -19.97 \\ -21.45 \\ -23.16 \\ \textbf{-25.06}
          \end{tabular}}  
          & \begin{tabular}[c]{@{}c@{}} 0.9960 \\ 0.9974  \\ 0.9985 \\ \textbf{0.9986}
          \end{tabular}  
          & \multicolumn{1}{c|}{ \begin{tabular}[c]{@{}c@{}} 1.184M \\ 1.183M \\ 626.576K \\ \textbf{86.4K} \end{tabular}}
          & \begin{tabular}[c]{@{}c@{}} \textbf{2.922M} \\ 3.869M \\ 27.018M \\ 24.762M 
          \end{tabular} 
          \\ \hline
    1/4  & \begin{tabular}[c]{@{}c@{}} CLNet\\ NR-CsiNet \\ FISTA-Net\\ MLCF \end{tabular}
          & \multicolumn{1}{c|}{\begin{tabular}[c]{@{}c@{}} -21.89 \\ -21.49 \\ -23.26 \\ \textbf{-30.40}
          \end{tabular}} 
          & \begin{tabular}[c]{@{}c@{}}  0.9977 \\ 0.9974 \\ 0.9985 \\  \textbf{0.9996}
          \end{tabular} 
          & \multicolumn{1}{c|}{ \begin{tabular}[c]{@{}c@{}} 2.364M \\ 2.363M \\ 1.216M \\ \textbf{96K} 
          \end{tabular}} 
          & \begin{tabular}[c]{@{}c@{}} \textbf{4.102M}  \\ 5.049M \\ 28.198M \\ 26.874M \end{tabular}
          \\ \hline
    \end{tabular}
\end{table}

Table.~\ref{tab2} summarizes the performance and complexity comparisons among the proposed MLCF and the benchmarks. With respect to the reconstruction performance, the best results for NMSE and cosine similarity $\rho$ are presented in bold fonts. The parameter $k$ is set to 50. One may observe that our proposed MLCF outperforms the other algorithms at all compression ratios. In particular, at a compression ratio of $\gamma$ = 1/4, our MLCF brings a gain of at least 7 dB in terms of the NMSE. 

Turning to complexity, the results of the lightest algorithm, i.e., the trainable parameters and floating point of operations (FLOPs), are shown in bold. From Table.~\ref{tab2}, it can be seen that the proposed MLCF has fewer trainable parameters. Coupled with the likewise small training data set, there is a relatively low training overhead for MLCF. However, the proposed MLCF is second only to FISTA-Net in FLOPs. MLCF sacrifices a certain amount of computational complexity in exchange for better reconstruction performance. In addition, it is noted that as the compression rate increases, the computational complexity of each method increases more.

\begin{figure}[htbp]
\centering {\includegraphics[width=\linewidth]{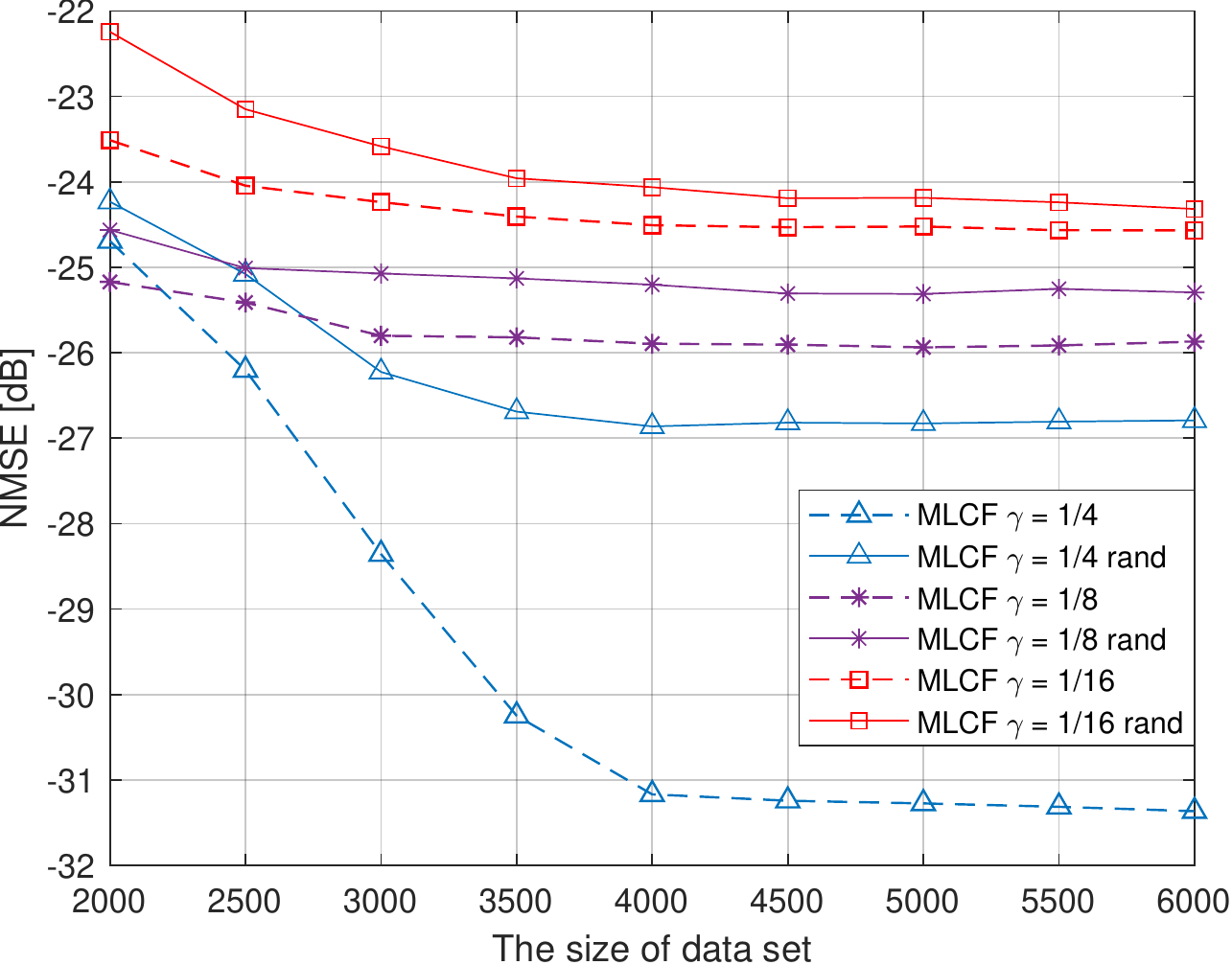}} 
\caption{The normalized mean square error vs. the size of data set ${\mathbf{X}}$.}
\label{fig2}
\end{figure}
Next, we show the impact of the size of data set ${\mathbf{X}}$ on the reconstruction performance, as shown in Fig.~\ref{fig2}. The values of $\lambda$ and $k$ are set to 0.05 and 80, respectively. For a certain compression ratio, the NMSE gradually decreases and eventually tends to converge as the size value $N$ increases, indicating that greater reconstruction performance may be obtained with a larger training size. Moreover, the computational complexity of Algorithm~\ref{alg1} is proportional to $N$, which determines the time required to learn the dictionaries. Based on this phenomenon, it is possible to make a trade-off between the reconstruction performance and the training time. In the meanwhile, we show the case of randomly chosen landmarks, labeled ``rand". The NMSE is lower when reconstructing the CSI with the proposed MLCF scheme, demonstrating the effectiveness of our scheme.

\begin{figure}[htbp]  
\centering {\includegraphics[width=\linewidth]{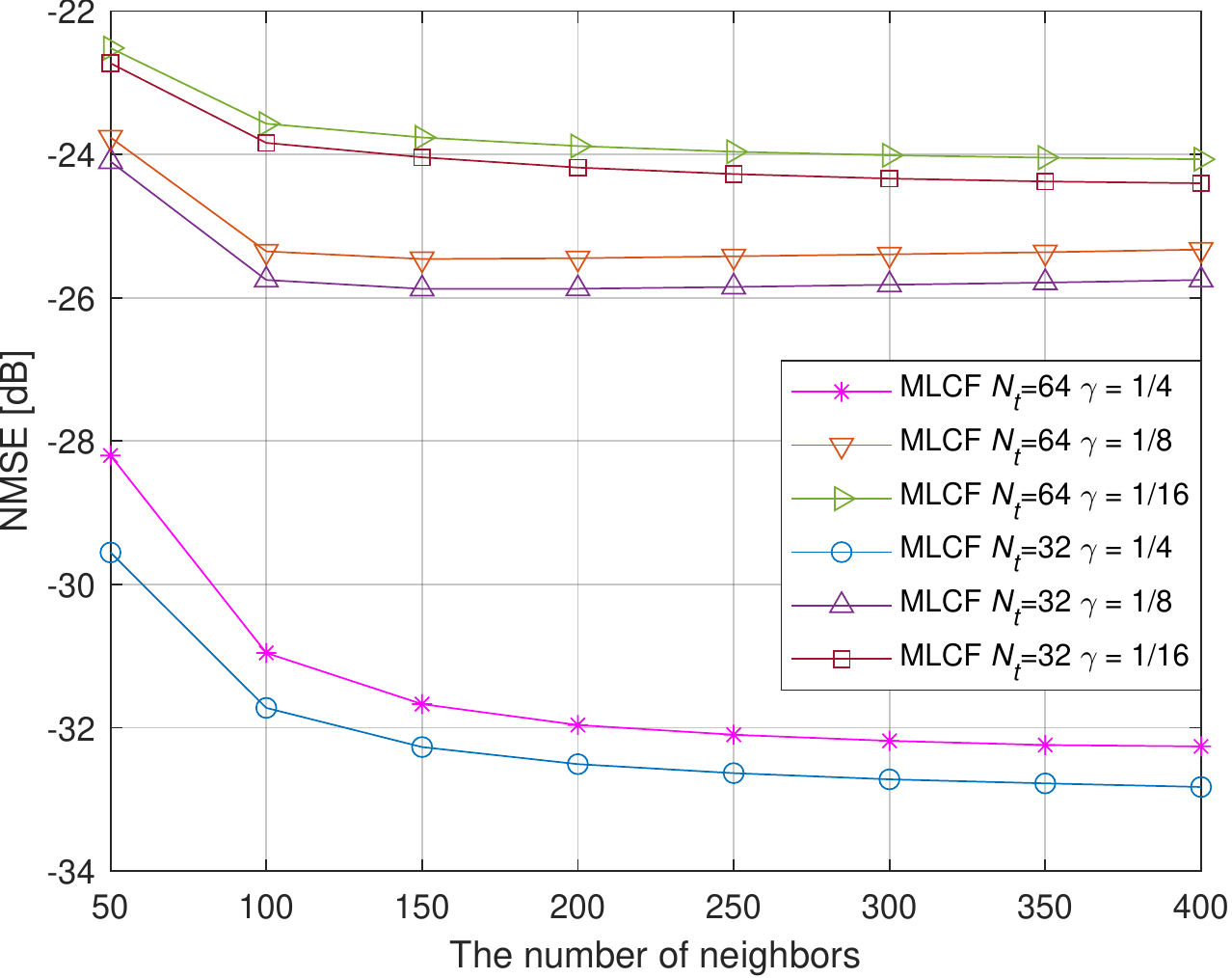}} 
\caption{The normalized mean square error vs. the number of neighbors $k$.}
\label{fig3}
\end{figure}
Fig.~\ref{fig3} depicts how the number of nearest neighbors $k$ affects the reconstruction quality NMSE. The size of dictionary $M$ is set to 600. The value $k$ ranges from 50 to 400 with a step size of 50. It can be found that as $k$ increases, the NMSE decreases, indicating that having more neighbors enhances the reconstruction performance. In fact, there is no systematic guide for the choice of $k$. Generally it can be selected based on Fig.~\ref{fig3} and adjusted according to the actual process. In addition, the BS antenna configurations of 32 and 64 are presented. For the same number of neighbors and compression ratio, the more antennas, the lower the reconstruction quality.

\begin{figure}[htbp]  
\centering {\includegraphics[width=\linewidth]{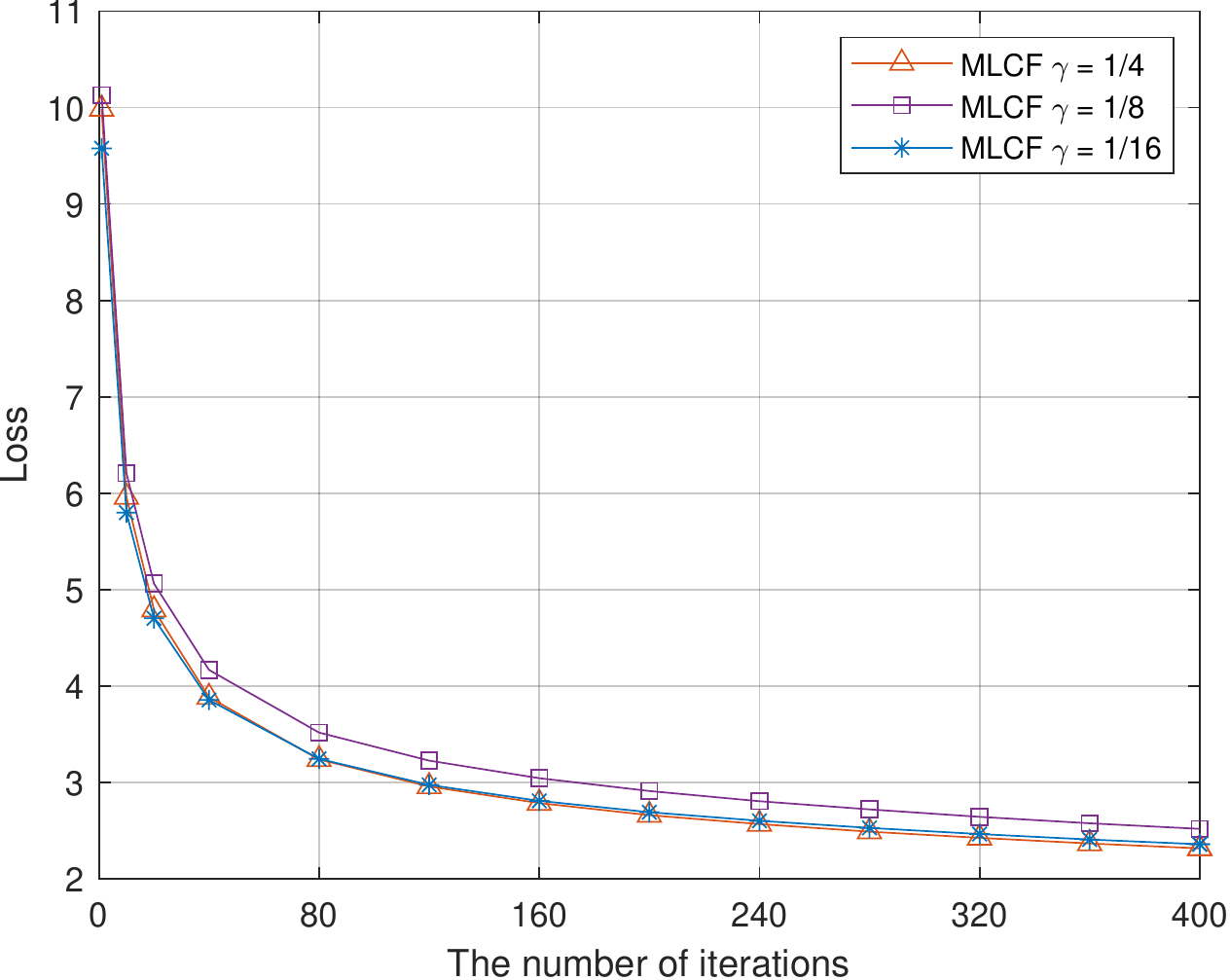}} 
\caption{The loss of the objective function Eq.~\eqref{final objective function} vs. the number of iterations.}
\label{fig4}
\end{figure}
Since the explicit mapping function $f$ is not available, the problem of dimensionality reduction is solved by optimizing the objective function Eq.~\eqref{final objective function}, which is the upper bound of the original objective function Eq.~\eqref{dr_objective function}. In this simulation, we verify the convergence of Algorithm~\ref{alg1} proposed to solve Eq.~\eqref{final objective function}, as shown in Fig.~\ref{fig4}. In each iteration, the updated dictionary and weight matrix are used to calculate the loss value of Eq.~\eqref{final objective function}. The loss value is affected by the training size $N$, thus it is divided by $N$ for normalization. We discover that the loss of Eq.~\eqref{final objective function} ultimately converges after numerous iterations, which is aligned with our theoretical analysis in Theorem~\ref{thm3}. 

\begin{figure}[htbp]  
\centering {\includegraphics[width=\linewidth]{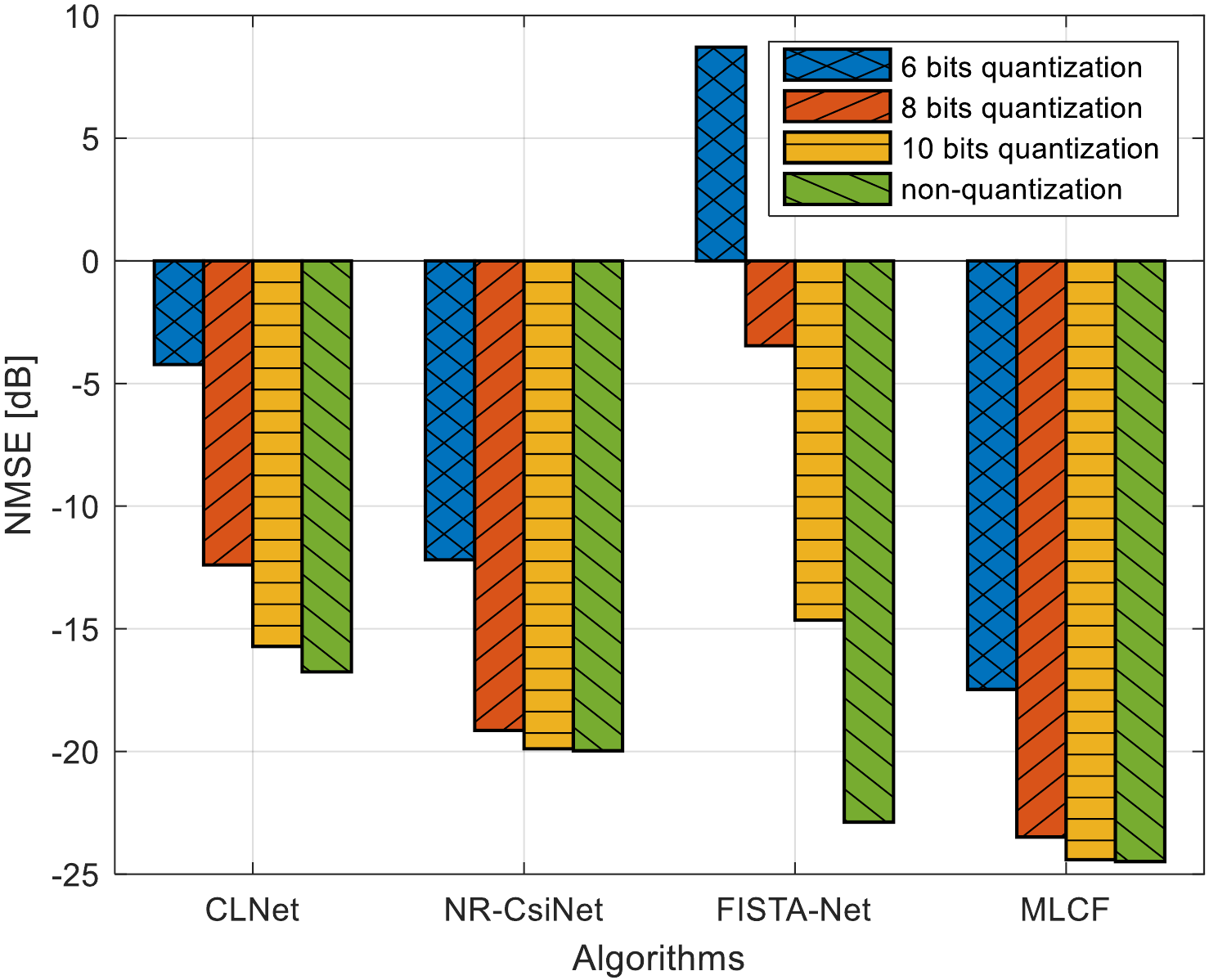}} 
\caption{The normalized mean square error vs. the algorithms with quantization and non-quantization.}
\label{fig5}
\end{figure}
To apply the limited feedback algorithms to the realistic communication scenarios, quantization is considered in the downlink CSI feedback process. That is, the compressed CSI is quantized by binary numbers before being transmitted to the BS. Uniform quantization is performed for the case of $\gamma$ = 1/16. The performance of NMSE under different quantization bits and different algorithms is shown in Fig.~\ref{fig5}. As can be observed, when the number of quantization bits is 10, the NMSE performance of all algorithms except FISTA-Net is not significantly worse than the unquantized situation. However, with lower quantization bits, a degradation in performance is observed, which is expected. The above results demonstrate that the proposed MLCF may be applicable in practical communication scenarios.

\begin{figure}[htbp]  
\centering {\includegraphics[width=\linewidth]{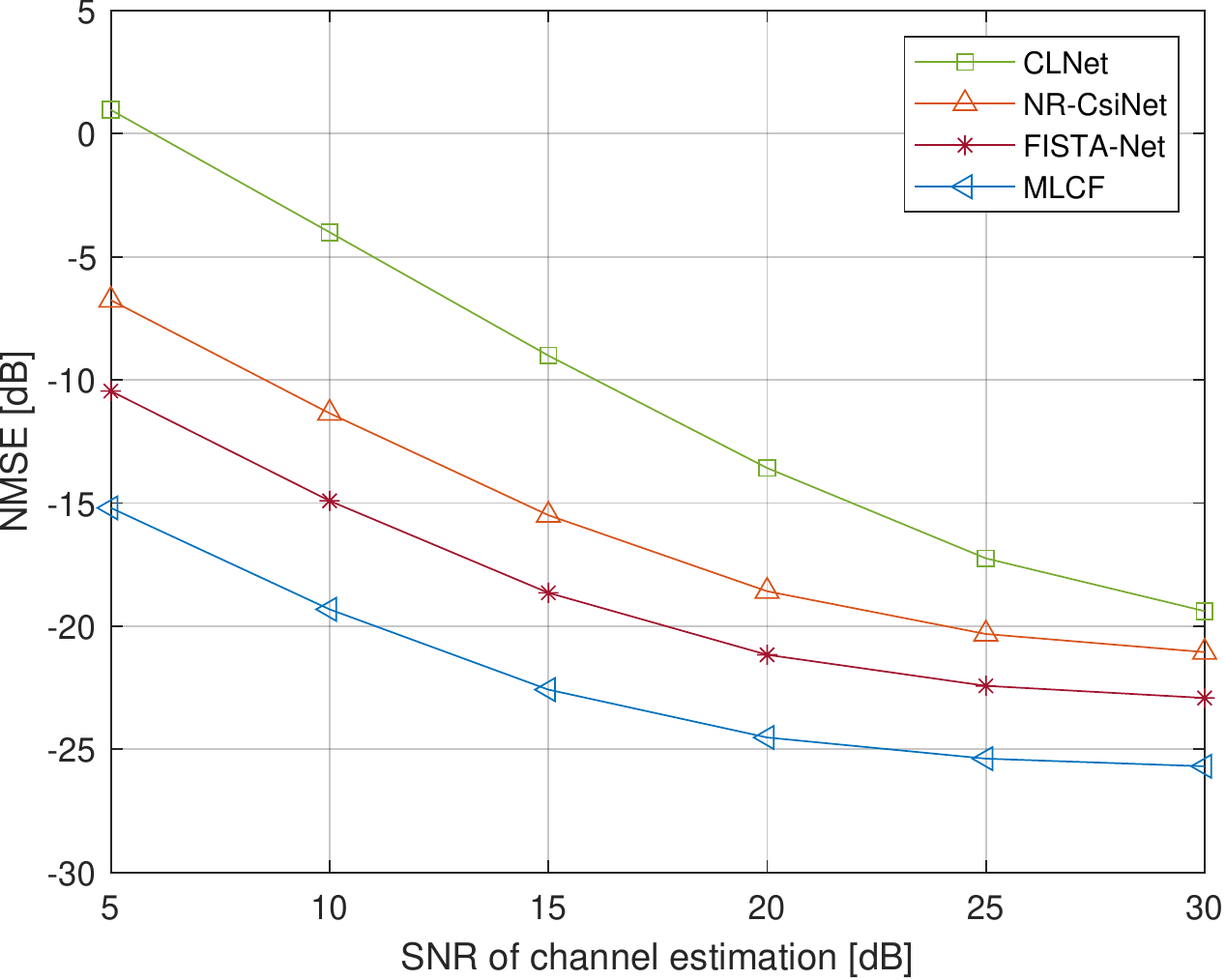}} 
\caption{The normalized mean square error vs. SNR of channel estimation.}
\label{fig6}
\end{figure}

We evaluate how the noise in the channel estimation affects the reconstruction quality. The networks of the benchmarks and the dictionaries of our MLCF are fixed in the simulation; they are all acquired through training with perfect CSI. The compression ratio is 1/8 and $\lambda$ is set to 0.05. And the signal-to-noise ratio (SNR) of channel estimation ranges from 5 dB to 30 dB. As shown in Fig.~\ref{fig6}, it is clear that our proposed MLCF performs well even under noisy channel estimation conditions, indicating that it is robust to noise. Furthermore, as the SNR of channel estimation rises, so does the reconstruction performance of each method.

\begin{figure}[htbp]  
\centering {\includegraphics[width=\linewidth]{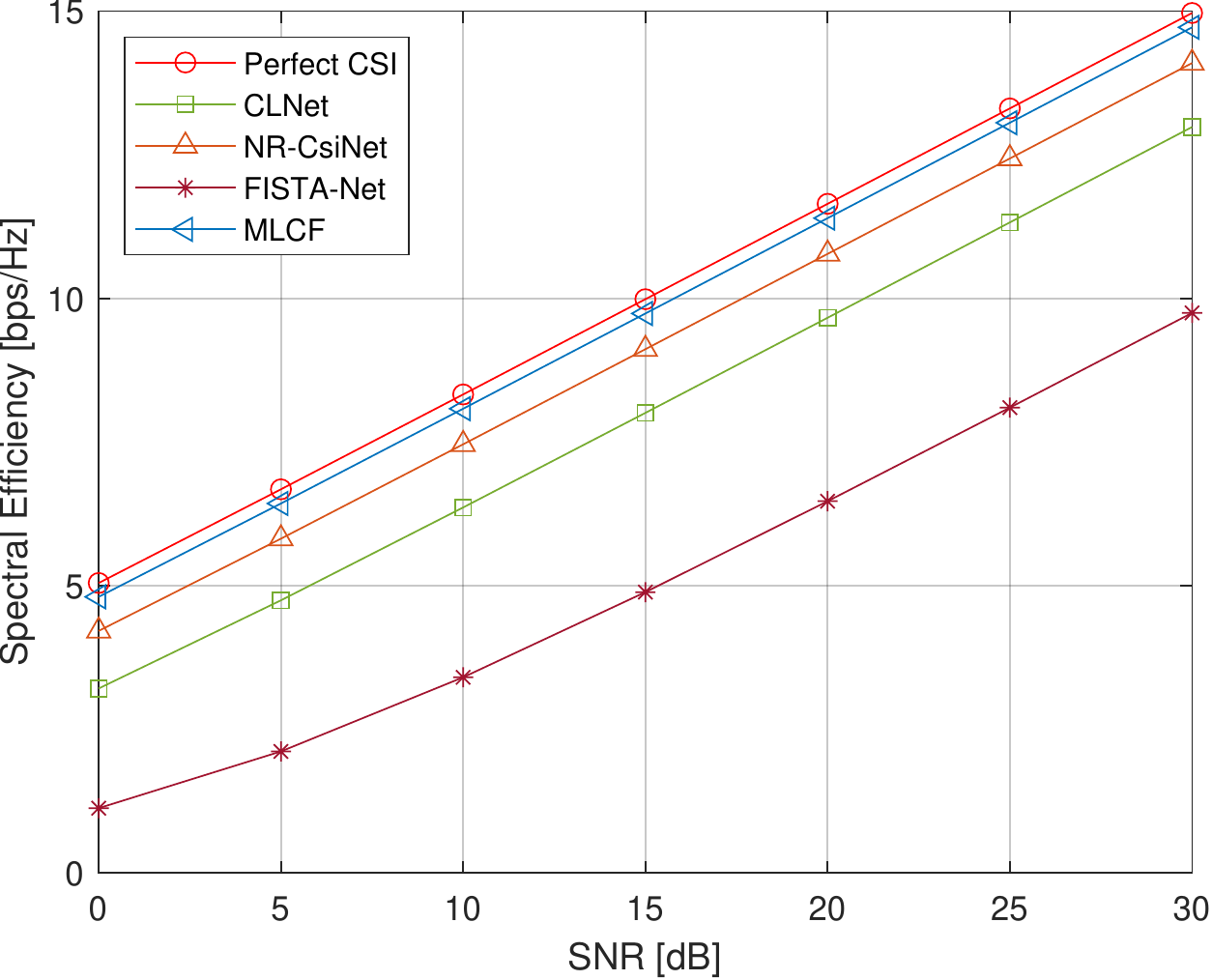}} 
\caption{The spectral efficiency vs. SNR.}
\label{fig7}
\end{figure}
Finally, Fig.~\ref{fig7} shows the spectral efficiency of FDD massive MIMO systems with different SNRs. The downlink precoder is the Eigen Zero-Forcing. The curve labeled ``Perfect CSI" means the BS has perfect downlink CSI, which is the upper-bound. The compression ratio and quantization bits are 1/16 and 4, respectively. The SE performance of our proposed MLCF is very close to the ideal case, indicating that the error between the reconstructed channel and the original channel is very small. 

\section{Conclusion}
In this paper, a novel manifold learning-based CSI feedback framework was proposed to reduce the feedback overhead of FDD massive MIMO systems. We considered the intrinsic manifold structure where the CSI samples reside. Without prior knowledge of the manifold structure, we constructed a data set consisting of high-density CSI samples to characterize it. In order to streamline the data set, a landmark selection algorithm was proposed to select the most representative landmarks so as to reconstruct all the CSI samples in the data set with minimum error. At the same time, the convergence of the algorithm was proved theoretically. Based on the pre-obtained landmarks, we proposed a low-complexity algorithm to efficiently compress and reconstruct the incremental CSI by keeping the local relationship with the landmarks unchanged. Simulation results showed that our proposed MLCF retained the advantages of manifold learning to achieve dimensionality reduction, and had superior reconstruction performance than existing DL-based algorithms.

\appendices
\section{Proof of Proposition~\ref{prop1}}
\label{appendixA}
\begin{prof} 
: For an arbitrary CSI sample ${{\mathbf{x}}_i} = g({{\mathbf{y}}_i})$, the linear structure of its  neighborhood can be characterized by the tangent space of the manifold $\mathcal{M}$ at ${{\mathbf{x}}_i}$. Assume the manifold is smooth enough and the dictionaries have been obtained. Using first-order Taylor expansion of $g$ at ${{\mathbf{x}}_i}$, a neighbor ${\mathbf{d}_j} = g({\mathbf{b}_j})$ in the landmarks can be represented as
\begin{equation}
    \label{tangent space}
    {\mathbf{d}_j} = {\mathbf{x}_i} + {\mathbf{J}_g}({\mathbf{y}_i}) ({\mathbf{b}_j} - {\mathbf{y}_i}) + {\boldsymbol{\delta}}({\mathbf{y}_i},{\mathbf{b}_j}) ,
\end{equation}
where ${\mathbf{J}_g}({\mathbf{y}_i}) \in  {\mathbb{R}^{2N_f \times d}}$ is the Jacobi matrix of $g$ at ${{\mathbf{y}}_i}$, whose column vectors span the tangent space, and ${\boldsymbol{\delta}}({\mathbf{y}_i},{{\mathbf{b}}_j})$ is the reminder term beyond the first-order expansion, which measures the approximation error of ${\mathbf{x}_i}$ to the tangent space. In particular, the $l$-th component of ${\boldsymbol{\delta}}({\mathbf{y}_i},{\mathbf{b}_j})$ is approximately equal to ${\delta}_l \approx  \frac{1}{2} {({\mathbf{b}_j} - {{\mathbf{y}}_i})^T}{\boldsymbol{\Psi}_{g_l}}({y_i})({\mathbf{b}_j} - {{\mathbf{y}}_i})$, where ${\boldsymbol{\Psi}_{g_l}}({{\mathbf{y}}_i})$ is the Hessian matrix of the $l$-th component function $g_l$ of $g$ at ${{\mathbf{y}}_i}$.

Based on the sum-to-one constraint $\sum\nolimits_{j = 1}^M {w_{ji}} = 1$, the error between the sample ${\mathbf{x}_i}$ and its local linear approximation is 
\begin{equation}
    \label{approximation error}
    \begin{aligned}
    {\varepsilon _i} = \left\| {{\mathbf{x}_i} - \sum\limits_{j = 1}^M {{w_{ji}}{\mathbf{d}_j}} } \right\|   
    = \left\| {\sum\limits_{j = 1}^M {w_{ji} \left( {{\mathbf{x}}_i} - {{\mathbf{d}}_j} \right)} } \right\|
    \end{aligned}. 
\end{equation}
Substituting Eq.~\eqref{tangent space} into the above error, we have
\begin{equation}
    \begin{aligned}
    {\varepsilon _i} &= \left\| {\sum\limits_{j = 1}^M {w_{ji}} \left( {{\mathbf{J}_g}({\mathbf{y}_i}) ({\mathbf{b}_i} - {\mathbf{y}_j}) + {\boldsymbol{\delta}}({\mathbf{y}_i},{\mathbf{b}_j})} \right)} \right\| \\
    & \le \left\| {{\mathbf{J}_g}({{\mathbf{y}}_i})\sum\limits_{j = 1}^M {{w_{ji}}({\mathbf{b}_i} - {\mathbf{y}_j})} } \right\| + \xi \sum\limits_{j = 1}^M \left\| {{\boldsymbol{\delta}}({\mathbf{y}_i},{\mathbf{b}_j})} \right\| \\
    & \le \left\| {{\mathbf{J}_g}({{\mathbf{y}}_i})\sum\limits_{j = 1}^M {{w_{ji}}({\mathbf{b}_i} - {\mathbf{y}_j})} } \right\| \\
    & \ \ \ \ \ \ \ \ \ \ \ \ + \xi \sum\limits_{j = 1}^M \sum\limits_{l = 1}^{2N_f} \left\| { { \frac{1}{2} {({\mathbf{b}_i} - {\mathbf{y}_j})}^T{\boldsymbol{\Psi}}_{g_l}({\mathbf{y}_i}) ({\mathbf{b}_i} - {{\mathbf{y}}_j})} } \right\| \\
    & \le \underbrace {\left\| {{\mathbf{J}_g}({{\mathbf{y}}_i} ) (\sum\limits_{j = 1}^M {w_{ji}}{\mathbf{b}_j} -{\mathbf{y}_i}) } \right\|}_{e_1} + \underbrace {{\xi_1}\left\| {\boldsymbol{\Psi}} \right\|_F \sum\limits_{j = 1}^M {{\left\| {\mathbf{b}_i} - {\mathbf{y}_j} \right\|}^2}}_{e_2} \\
    & \le \underbrace{{\xi} \left\| {\mathbf{J}_g}({\mathbf{y}_i}) \right\|_F  \sum\limits_{j = 1}^M  \left\| {\mathbf{b}_j} - {{\mathbf{y}}_i}\right\|}_{e_1} + \underbrace{{\xi_1}\left\| {\boldsymbol{\Psi}} \right\|_F \sum\limits_{j = 1}^M {{\left\| {\mathbf{b}_j} - {{\mathbf{y}}_i} \right\|}^2}}_{e_2} , 
    \end{aligned}
\end{equation}
where $\xi = \max {\{ {w_{ji}}\} }_j$ is the largest entry in ${{\mathbf{w}}_i}$, $\xi_1 = \sqrt{\frac{N_f}{2} }\xi$, and ${\boldsymbol{\Psi}} = \left[ {\begin{array}{*{20}{c}} {\boldsymbol{\Psi}^T_{g_1}}({{\mathbf{y}}_i})& \ldots &{\boldsymbol{\Psi}^T_{g_{2N_f}}}({{\mathbf{y}}_i}) \end{array}} \right]^T \in  {\mathbb{R}^{2N_fd \times d}} $.

Two terms make up the approximation error, as can be seen. In the first term, since the column vectors of ${\mathbf{J}_g}({\mathbf{y}_i})$ span the tangent space of $g$ at ${\mathbf{y}_i}$, ${\mathbf{J}_g}({\mathbf{y}_i}){\mathbf{y}_i}$ represents the projection of ${\mathbf{y}_i}$ on the tangent space. Therefore, $e_1$ reflects the projected distance between ${\mathbf{y}_i}$ and $\sum\nolimits_j {{w_{ji}}{\mathbf{b}_j}} $ on the tangent space. In the second term, the Hessian matrix ${\boldsymbol{\Psi}}$ and its upper bound ${\left\| {\boldsymbol{\Psi}} \right\|_F}$ are determined by the local curvature of the manifold at ${{\mathbf{x}}_i}$. We have no prior knowledge of the manifold structure in a learning task, hence the constrains on the manifold such as local curvature are impractical in real implementations. 

Taking another shortcut, we consider the influence of neighbors on the approximation error. If the neighbors of ${\mathbf{x}_i}$ in the dictionary ${\mathbf{D}}_{\rm{H}}^{\rm{dr}}$ lie in a sufficiently compact region, the hyperplane spanned by the neighbors is almost the same as the tangent space. In this case, $\max {\{ \left\| {\mathbf{b}_j} - {\mathbf{y}_i}\right\| \}}_j$ will be tiny. Meanwhile, the local curvature at ${\mathbf{x}_i}$ and ${\left\| {\boldsymbol{\Psi}} \right\|}_F$ will be zero or close to zero. At this point, the approximation error ${\varepsilon _i}$ is undoubtedly quite small. Thus, Proposition~\ref{prop1} is proved.
\end{prof}

\section{Proof of Theorem~\ref{thm1}}
\label{appendixB}
\begin{prof}
: Eliminating the terms irrelevant to ${\mathbf{d}}_j$, Eq.~\eqref{dr_high dictionary objective function} can be simplified to 
\begin{equation}
    \label{dictionary}
    \begin{aligned}
    \mathcal{L}({\mathbf{d}}_j) & = \| {{\mathbf{E}} - {{\mathbf{d}}_j}{{\mathbf{W}}_{j*}} } \|_F^2 + \lambda \sum\limits_{i = 1}^N {w_{ji}^2{{\| {{{\mathbf{x}}_i} - {{\mathbf{d}}_j}} \|}^2}}  \\ 
    & = {\rm{tr}} \left\{ ({\mathbf{E}}  - {{\mathbf{d}}_j}{{\mathbf{W}}_{j*}} ) ({\mathbf{E}} - {{\mathbf{d}}_j}{{\mathbf{W}}_{j*}} )^T \right\} \\
    & \ \ \ \ \ \ \ \ \ \ \  + \lambda \sum\limits_{i = 1}^N {w_{ji}^2 ({{\mathbf{x}}_i} - {{\mathbf{d}}_j})^T ({{\mathbf{x}}_i} - {{\mathbf{d}}_j})}
    \end{aligned},
\end{equation}
where each term employs the $L_p$-norm regularizer and 
satisfies $ p \ge 1$. Obviously, the objective function Eq.~\eqref{dictionary} is convex. 

The gradient of $\mathcal{L}({\mathbf{d}}_j)$ with regard to ${{\mathbf{d}}_j}$ is 
\begin{equation}
    \frac {\partial \mathcal{L}({\mathbf{d}}_j)}  {\partial {\mathbf{d}}_j} = -2({\mathbf{E}}  - {{\mathbf{d}}_j}{{\mathbf{W}}_{j*}}){{\mathbf{W}}_{j*}^T} - 2 \lambda \sum\limits_{i = 1}^N {w_{ji}^2} ({{\mathbf{x}}_i} - {{\mathbf{d}}_j}) .
\end{equation}
By setting ${{\partial \mathcal{L}({\mathbf{d}}_j)} \mathord{\left/
 {\vphantom {{\partial \mathcal{L}({\mathbf{d}}_j)} {\partial {\mathbf{d}_j}}}} \right.
 \kern-\nulldelimiterspace} {\partial {\mathbf{d}}_j}}$ to be zero, the optimal solution to ${\mathbf{d}_j}$ is as follows
 \begin{equation}
    {{\mathbf{d}}_j} = \frac{{\mathbf{E}} {\mathbf{W}}_{j*}^T + \lambda {\mathbf{X}}{({\mathbf{W}_{j*}^2)}^T}} {( 1 + \lambda ) {{{\mathbf{W}}_{j*}} {{\mathbf{W}}}_{j*}^T} },
\end{equation}
where ${\mathbf{W}}_{j*}^2$ represents the square of each entry in ${\mathbf{W}}_{j*}$, and ${\mathbf{X}}{({\mathbf{W}_{j*}^2})}^T = \sum\limits_{i = 1}^N w_{ji}^2 {{\mathbf{x}}_i}$. Thus, Theorem~\ref{thm1} is proved.
\end{prof}

\section{Proof of Theorem~\ref{thm2}}
\label{appendixC}
\begin{prof}
: Taking the constraint ${\mathbf{e}}^T {\hat {\mathbf{w}}_i} = 1$ into account, the first term in Eq.~\eqref{dr_coding objective function} can be rewritten as
\begin{equation}
     {\left\| {{\mathbf{x}}_i} - {\mathbf{N}}_{{\mathbf{x}}_i}{{\hat {\mathbf{w}}}_i} \right\|^2}  = {\left\| ( {\mathbf{x}_i}{\mathbf{e}}^T  - {\mathbf{N}}_{{\mathbf{x}}_i}) {\hat {\mathbf{w}}_i}  \right\|^2} 
    = {{\hat {\mathbf{w}}}_i^T} {\mathbf{R}} {{\hat {\mathbf{w}}}_i} ,
\end{equation}
where ${\mathbf{R}} = ({\mathbf{x}_i}{\mathbf{e}}^T  - {\mathbf{N}}_{{\mathbf{x}}_i})^T ({\mathbf{x}_i}{\mathbf{e}}^T - {\mathbf{N}}_{{\mathbf{x}}_i})$. 

The second term can be reformulated as
\begin{equation}
    \sum\limits_{j = 1}^k {\hat w_{ji}^2{{\| {\mathbf{x}_i} - {\mathbf{d}_{a_j}} \|}^2}} = {{\hat {\mathbf{w}}}_i^T} \varphi({\mathbf{R}})  {{\hat {\mathbf{w}}}_i} .
\end{equation}
Note that ${\|{\mathbf{x}_i} - {\mathbf{d}_{a_1}} \|}^2, \ldots , {\| {\mathbf{x}_i} - {\mathbf{d}_{a_k}} \|}^2$ are the same as the diagonal elements of ${\mathbf{R}}$. We define a matrix operator $\varphi(\cdot)$ that preserves only the diagonal elements of the matrix and sets the rest to zero. Therefore, for simplicity, we have $\varphi({\mathbf{R}}) = {\rm{diag}} ({{\| {\mathbf{x}_i} - {\mathbf{d}_{a_1}} \|}^2}, \ldots , {{\| {\mathbf{x}_i} - {\mathbf{d}_{a_k}} \|}^2} )$.

The derivative of the third term ${\left\| {\mathbf{w}} \right\|_{2,1}}$ equals the derivative of ${\rm{tr}}({\mathbf{w}}^T{\mathbf{G}}{\mathbf{w}})$, where ${\mathbf{G}}$ has been defined in Eq.~\eqref{g}.

The above three items all satisfy the norm greater than or equal to 1, thus it is clear that Eq.~\eqref{dr_coding objective function} is a convex function. Introducing the Lagrange multiplier, the optimization problem can be reformulated as
\begin{equation}
    \begin{aligned}
    \mathcal{L}({{\hat {\mathbf{w}}}_i}, \eta) & = {\left\| {{\mathbf{x}}_i} - {\mathbf{N}}_{{\mathbf{x}}_i}{{\hat {\mathbf{w}}}_i} \right\|^2} + \lambda \sum\limits_{j = 1}^k {\hat w_{ji}^2{{\| {\mathbf{x}_i} - {\mathbf{d}_{a_j}} \|}^2}}  \\ & \ \ \  + \mu {\rm{tr}}({\mathbf{W}}^T{\mathbf{G}}{\mathbf{W}}) + \eta({\mathbf{e}}^T{{\hat {\mathbf{w}}}_i} - 1) \\
    & = {{\hat {\mathbf{w}}}_i^T} \big({\mathbf{R}} + \lambda\varphi(\mathbf{R})\big) {{\hat {\mathbf{w}}}_i} + \mu {\rm{tr}}({\mathbf{W}}^T{\mathbf{G}}{\mathbf{W}})  \\
    & \ \ \  + \eta({\mathbf{e}}^T{{\hat {\mathbf{w}}}_i} - 1) 
    \end{aligned} .
\end{equation}
Setting the partial derivatives of $\mathcal{L}({{\hat {\mathbf{w}}}_i}, \eta)$ with regard to ${{\hat {\mathbf{w}}}_i}$ and $\eta$ to be zero, we have
\begin{equation}
    \left\{ { \begin{aligned}
    & {\frac{{\partial {\cal \mathcal{L}}}}{{\partial {{\hat {\mathbf{w}}}_i}}} = 2 \left({\mathbf{R}} + \lambda \varphi ({\mathbf{R}}) \right){{\hat {\mathbf{w}}}_i} + 2 \mu {\hat{ \mathbf{G}}}{{\hat {\mathbf{w}}}_i} + \eta {\mathbf{e}} = 0}\\
    &\ {\frac{{\partial {\cal \mathcal{L}}}}{{\partial \eta }} = {{\mathbf{e}}^T}{{\hat {\mathbf{w}}}_i} - 1 = 0}
    \end{aligned} } \right. ,
\end{equation}
where ${\hat{ \mathbf{G}}} = {\rm{diag}}(g_{a_1 a_1},g_{a_2 a_2},\ldots,g_{a_k a_k}) \in {\mathbb{R}^{k \times k}}$. Therefore, we have the optimal solution to the weight sub-vector as
\begin{equation}
    {\hat {\mathbf{w}}_i} =  \frac{{{\left( {\mathbf{R}} + \lambda \varphi ({\mathbf{R}}) + \mu{\hat{ \mathbf{G}}} \right)}^{ - 1}}{\mathbf{e}}} {{\mathbf{e}^T}{\left( {\mathbf{R}} + \lambda \varphi ({\mathbf{R}}) +\mu{\hat{ \mathbf{G}}} \right)^{ - 1}}{\mathbf{e}}} .
\end{equation}
Thus, Theorem~\ref{thm2} is proved.
\end{prof}

\ifCLASSOPTIONcaptionsoff
  \newpage
\fi

\bibliographystyle{IEEEtran}
\bibliography{IEEEabrv,reference}

\begin{IEEEbiography}
[{\includegraphics[width=1in,height=1.25in,clip,keepaspectratio]{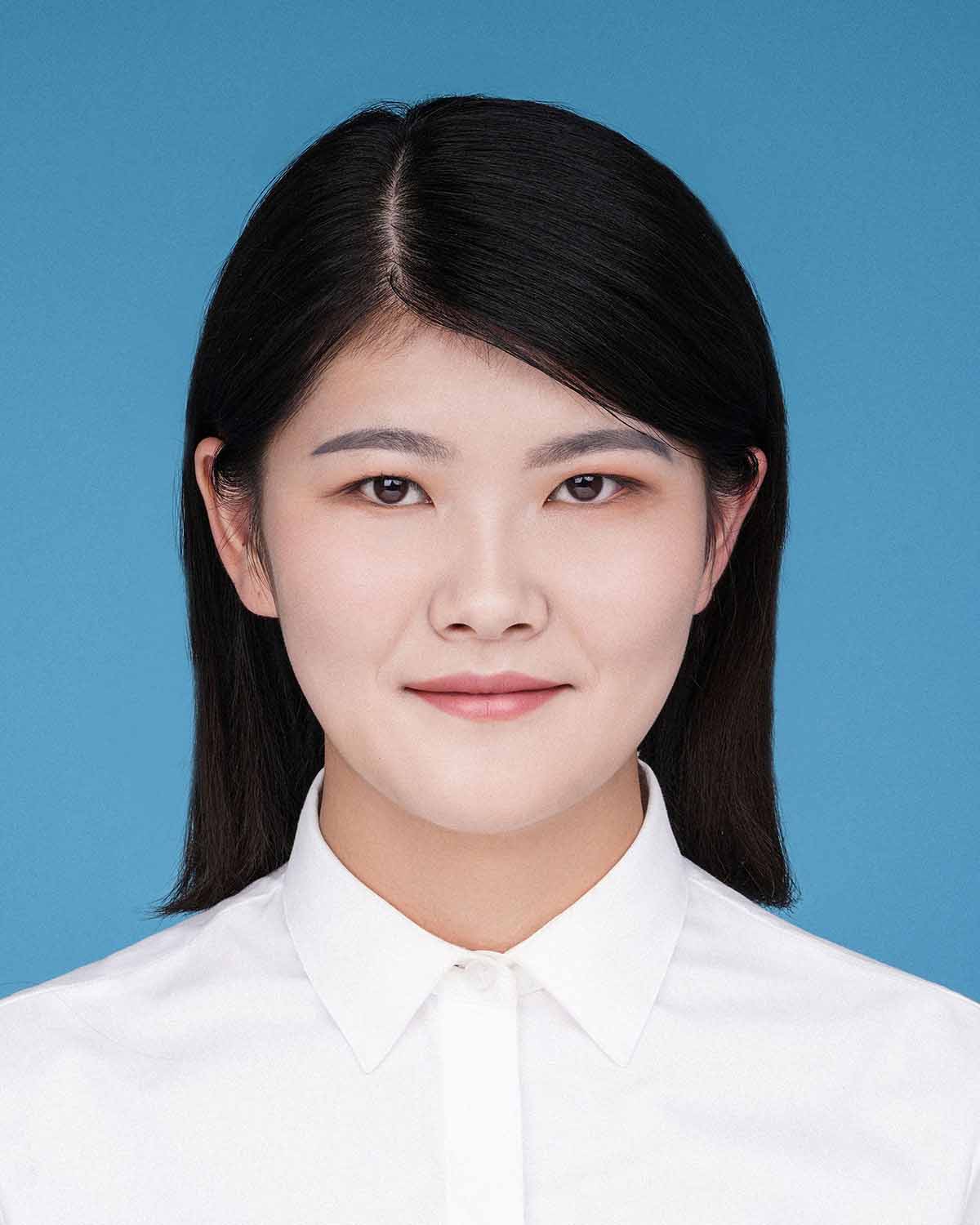}}]{Yandi Cao}
received the B.Sc. degree in Communication Engineering from the School of Microelectronics and Communication Engineering, Chongqing University, Chongqing, China, in 2020. She is currently pursuing the Ph.D. degree with the School of Electronic Information and Communications from Huazhong University of Science and Technology, Wuhan, China. Her research interests include machine learning, signal processing, and channel feedback for massive MIMO systems.
\end{IEEEbiography}

\begin{IEEEbiography}[{\includegraphics[width=1in,height=1.25in,clip,keepaspectratio]{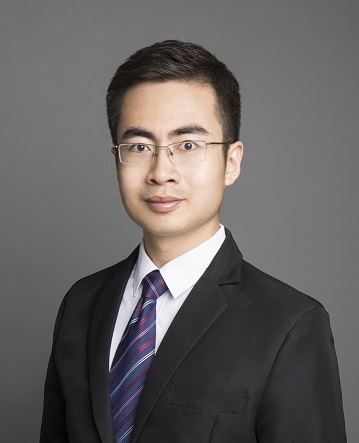}}]{Haifan Yin}
(Senior Member, IEEE) received the B.Sc. degree in electrical and electronic engineering and the M.Sc. degree in electronics and information engineering from the Huazhong University of Science and Technology, Wuhan, China, in 2009 and 2012, respectively, and the Ph.D. degree from Télécom ParisTech in 2015. From 2009 to 2011, he was a Research and Development Engineer with the Wuhan National Laboratory for Optoelectronics, Wuhan, working on the implementation of TD-LTE systems. From 2016 to 2017, he was a DSP Engineer at Sequans Communications (IoT chipmaker), Paris, France. From 2017 to 2019, he was a Senior Research Engineer working on 5G standardization at Shanghai Huawei Technologies Company Ltd., where he has made substantial contributions to 5G standards, particularly the 5G codebooks. Since May 2019, he has been a Full Professor with the School of Electronic Information and Communications, Huazhong University of Science and Technology. His current research interests include 5G and 6G networks, signal processing, machine learning, and massive MIMO systems. He was the National Champion of 2021 High Potential Innovation Prize awarded by the Chinese Academy of Engineering, a recipient of the China Youth May Fourth Medal (the top honor for young Chinese), and a recipient of the 2024 Stephen O. Rice Prize.
\end{IEEEbiography}

\begin{IEEEbiography}
[{\includegraphics[width=1in,height=1.25in,clip,keepaspectratio]{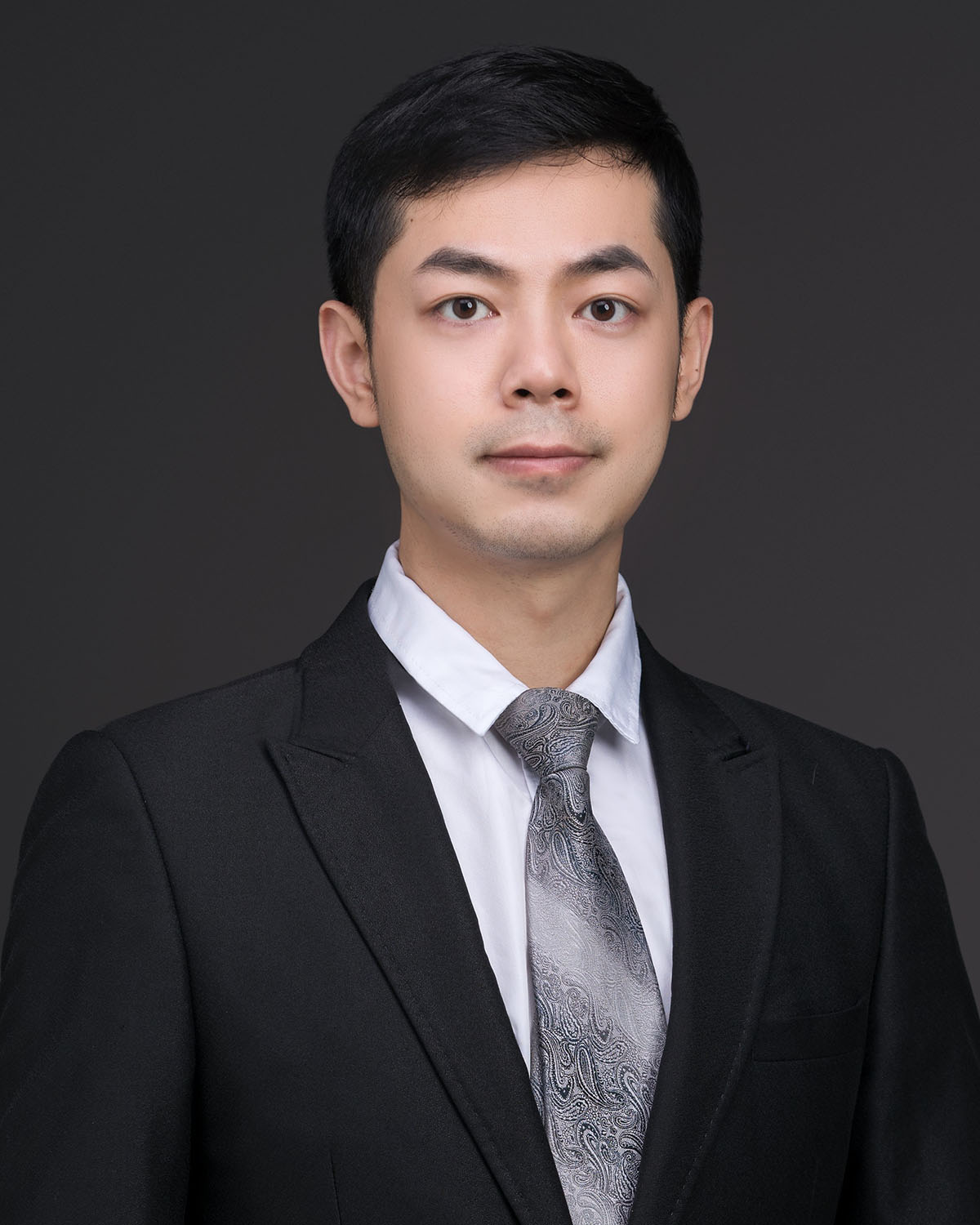}}]{Ziao Qin}
received the B.S. degree in information engineering from Beijing Institute of Technology, Beijing, China, in 2014. From 2014 to 2017, he worked in industry in Beijing, China. He received the Ph.D. degree in information and communications engineering from Huazhong University of Science and Technology, Wuhan, China in 2024. His research interests include channel estimation, signal processing, codebook design and beamforming for massive MIMO systems. 
\end{IEEEbiography}

\begin{IEEEbiography}[{\includegraphics[width=1in,height=1.25in,clip,keepaspectratio]{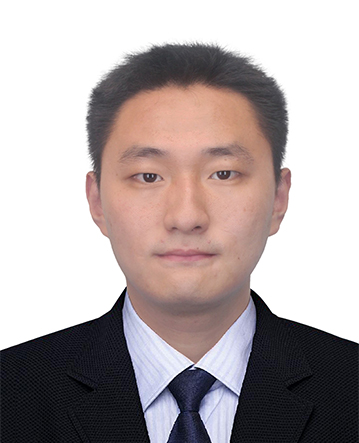}}]{Weidong Li}
received the B.Sc degree in electronic information science and technology from Nanjing Agricultural University, Nanjing, China, in 2017, and the M.Sc degree in electronic engineering from the Nanjing University of Aeronautics and Astronautics, Nanjing, China, in 2020. He is currently pursuing the Ph.D. degree with the School of Electronic Information and Communications from Huazhong University of Science and Technology, Wuhan, China. His research interests include channel estimation, signal processing, and the mobility of massive MIMO and ELAA.
\end{IEEEbiography}

\begin{IEEEbiography}[{\includegraphics[width=1in,height=1.25in,clip,keepaspectratio]{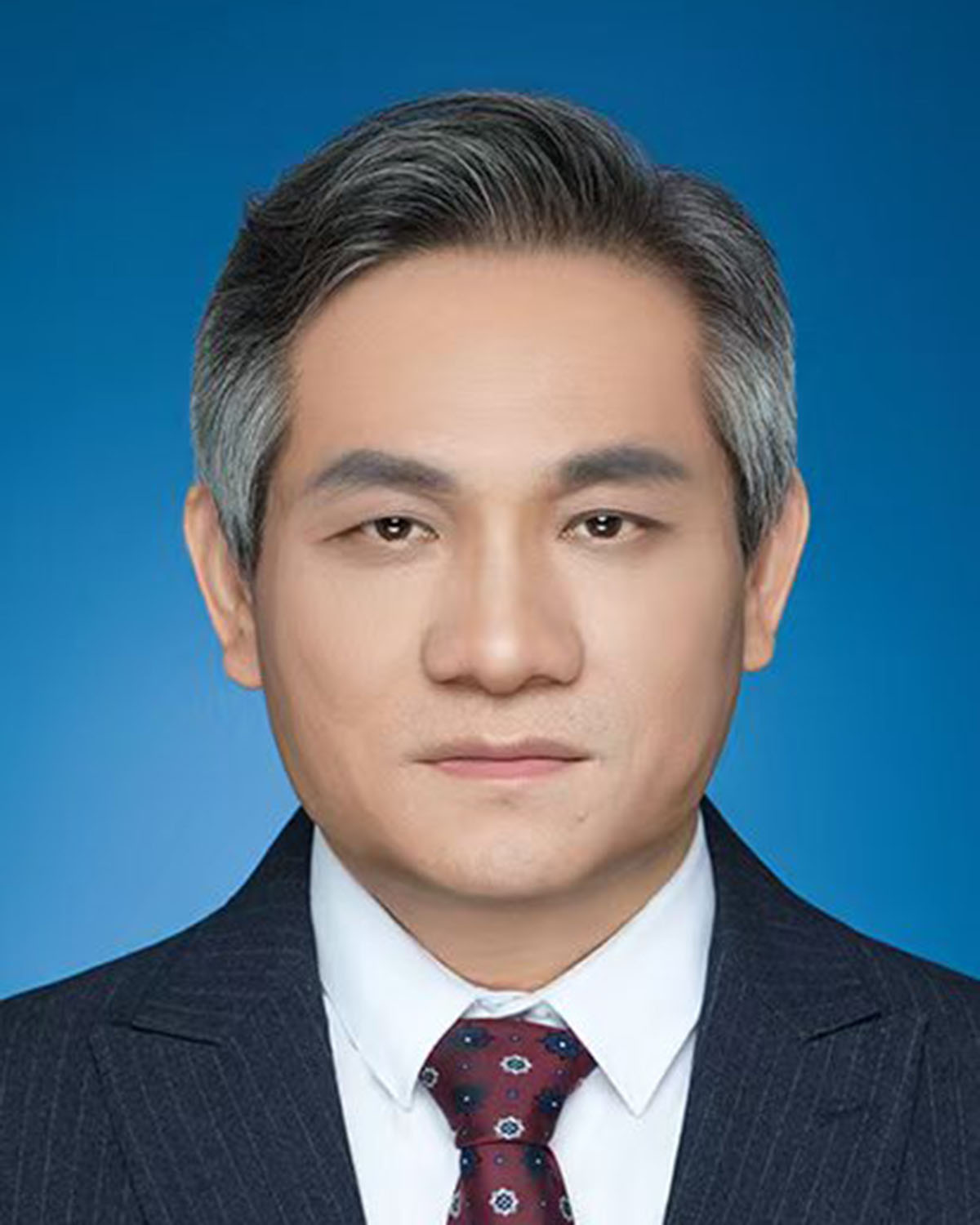}}]{Weimin Wu}
received the B.E. degree in computer software from Xidian University, Xi'an, China, in 1992, the M.E. degree in computer application from Sichuan University, Chengdu, China, in 1995, and the Ph.D. degree in communications and information systems from the Huazhong University of Science and Technology, Wuhan, China, in 2007. He is currently an Associate Professor with the School of Electronic Information and Communications, Huazhong University of Science and Technology. His current research interests include Internet streaming, broadband wireless communications, and networking.
\end{IEEEbiography}

\begin{IEEEbiography}
[{\includegraphics[width=1in,height=1.25in,clip,keepaspectratio]{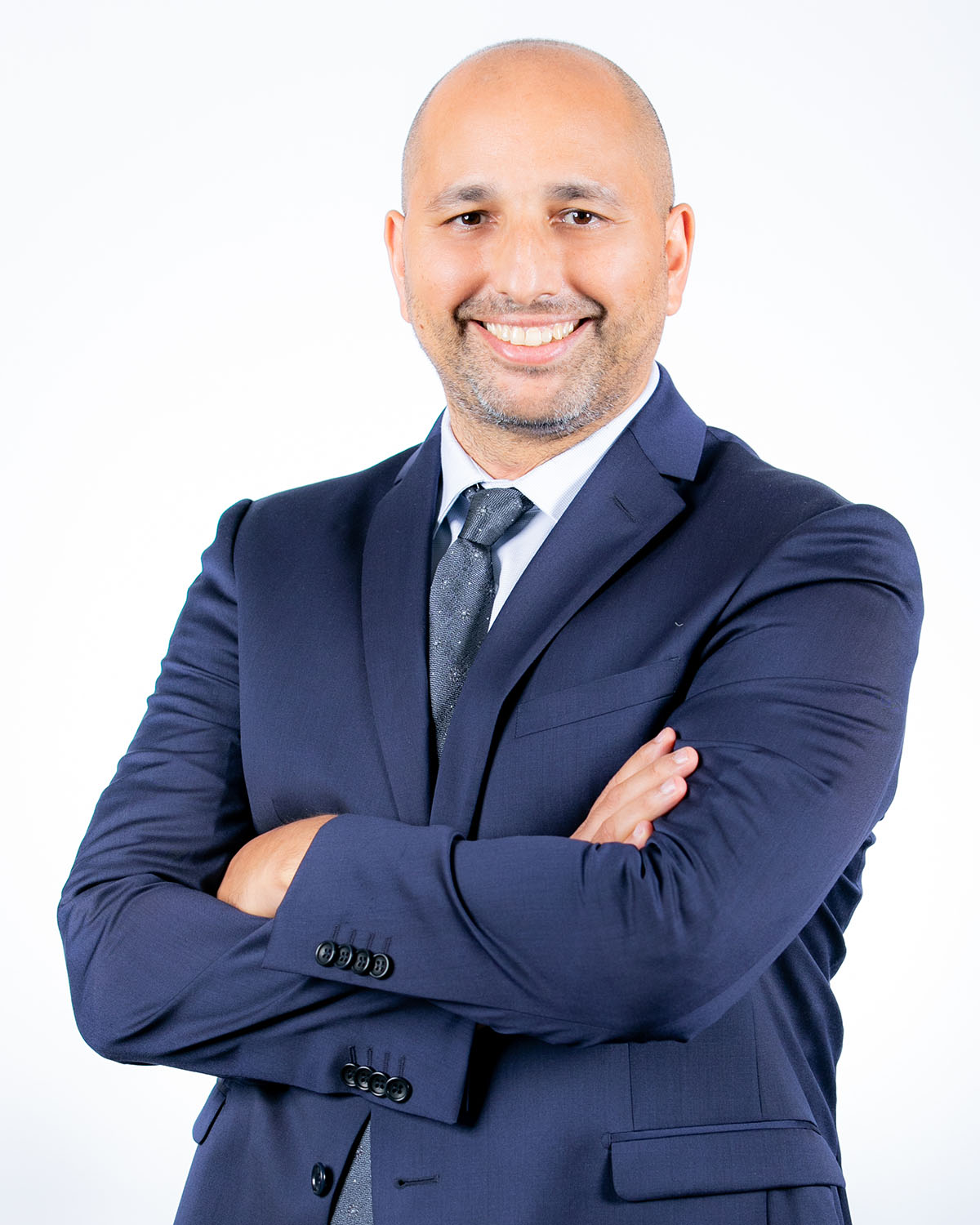}}]{M\'{e}rouane Debbah}
(Fellow, IEEE) is Chief Researcher at the Technology Innovation Institute in Abu Dhabi. He is a Professor at Centralesupelec and an Adjunct Professor with the Department of Machine Learning at the Mohamed Bin Zayed University of Artificial Intelligence. He received the M.Sc. and Ph.D. degrees from the Ecole Normale Supérieure Paris-Saclay, France. He was with Motorola Labs, Saclay, France, from 1999 to 2002, and also with the Vienna Research Center for Telecommunications, Vienna, Austria, until 2003. From 2003 to 2007, he was an Assistant Professor with the Mobile Communications Department, Institut Eurecom, Sophia Antipolis, France. In 2007, he was appointed Full Professor at CentraleSupelec, Gif-sur-Yvette, France. From 2007 to 2014, he was the Director of the Alcatel-Lucent Chair on Flexible Radio. From 2014 to 2021, he was Vice-President of the Huawei France Research Center. He was jointly the director of the Mathematical and Algorithmic Sciences Lab as well as the director of the Lagrange Mathematical and Computing Research Center. Since 2021, he is leading the AI \& Digital Science Research centers at the Technology Innovation Institute. He has managed 8 EU projects and more than 24 national and international projects. His research interests lie in fundamental mathematics, algorithms, statistics, information, and communication sciences research. He is an IEEE Fellow, a WWRF Fellow, a Eurasip Fellow, an AAIA Fellow, an Institut Louis Bachelier Fellow and a Membre émérite SEE. He was a recipient of the ERC Grant MORE (Advanced Mathematical Tools for Complex Network Engineering) from 2012 to 2017. He was a recipient of the Mario Boella Award in 2005, the IEEE Glavieux Prize Award in 2011, the Qualcomm Innovation Prize Award in 2012, the 2019 IEEE Radio Communications Committee Technical Recognition Award and the 2020 SEE Blondel Medal. He received more than 20 best paper awards, among which the 2007 IEEE GLOBECOM Best Paper Award, the Wi-Opt 2009 Best Paper Award, the 2010 Newcom++ Best Paper Award, the WUN CogCom Best Paper 2012 and 2013 Award, the 2014 WCNC Best Paper Award, the 2015 ICC Best Paper Award, the 2015 IEEE Communications Society Leonard G. Abraham Prize, the 2015 IEEE Communications Society Fred W. Ellersick Prize, the 2016 IEEE Communications Society Best Tutorial Paper Award, the 2016 European Wireless Best Paper Award, the 2017 Eurasip Best Paper Award, the 2018 IEEE Marconi Prize Paper Award, the 2019 IEEE Communications Society Young Author Best Paper Award, the 2021 Eurasip Best Paper Award, the 2021 IEEE Marconi Prize Paper Award, the 2022 IEEE Communications Society Outstanding Paper Award, the 2022  ICC Best paper Award as well as the Valuetools 2007, Valuetools 2008, CrownCom 2009, Valuetools 2012, SAM 2014, and 2017 IEEE Sweden VT-COM-IT Joint Chapter best student paper awards. He is an Associate Editor-in-Chief of the journal Random Matrix: Theory and Applications. He was an Associate Area Editor and Senior Area Editor of the IEEE TRANSACTIONS ON SIGNAL PROCESSING from 2011 to 2013 and from 2013 to 2014, respectively. From 2021 to 2022, he serves as an IEEE Signal Processing Society Distinguished Industry Speaker. 
\end{IEEEbiography}

\end{document}